\documentclass[lettersize,journal]{IEEEtran}

\usepackage{amsmath,amsfonts,amssymb}
\DeclareMathOperator{\sinc}{sinc}

\usepackage{graphicx}
\usepackage[caption=false,font=footnotesize]{subfig}
\usepackage{array}
\usepackage{tabularx}
\usepackage{makecell}
\usepackage{multirow}
\usepackage{threeparttable}
\usepackage{stfloats}

\usepackage{algorithm}
\usepackage{algorithmic}

\usepackage{cite}
\usepackage{url}
\usepackage{textcomp}
\usepackage{verbatim}
\usepackage{color}

\usepackage{physics}

\begin{document}

\title{Spatial Correlation, Non-Stationarity, and Degrees of Freedom of Holographic Curvature-Reconfigurable Apertures}

\author{Liuxun~Xue,~\IEEEmembership{Graduate Student Member,~IEEE, }%
        Shu~Sun,~\IEEEmembership{Senior Member,~IEEE, }%
        Ruifeng~Gao,~\IEEEmembership{Member,~IEEE, }%
        and Xiaoqian~Yi,~\IEEEmembership{Senior Member,~IEEE}%
\thanks{
A preliminary conference version of part of this work appeared in~\cite{VTC2025}. The present manuscript substantially extends that work with a visibility-aware spatial characterization framework for HoloCuRA, including new spatial non-stationarity analysis, additional propagation scenarios, and new theoretical and numerical results.
}%
\thanks{Liuxun Xue and Shu Sun are with the School of Information and Electronic Engineering,
Shanghai Jiao Tong University, Shanghai 200240, China. (e-mail: liuxun66@sjtu.edu.cn; shusun@sjtu.edu.cn).

Ruifeng Gao is with the School of Transportation and Civil Engineering, Nantong University, Nantong 226019, China (e-mail: grf@ntu.edu.cn).

Xiaoqian Yi is with the Flight Test Center, Commercial Aircraft Corporation of China (COMAC), Shanghai 201323, China (e-mail: yixiaoqian@comac.cc).}%
}

\maketitle

\begin{abstract}
Low-altitude wireless platforms increasingly require lightweight, conformal, and densely sampled antenna array apertures with high array gain and spatial selectivity. 
However, when deployed on nonplanar surfaces, curvature alters the array manifold, local visibility, and propagation support, potentially invalidating spatial-stationarity assumptions.
In this paper, we investigate a holographic curvature-reconfigurable aperture (HoloCuRA), modeled as a curvature-controllable holographic surface, and develop a visibility-aware spatial characterization framework for its low-altitude applications. 
Specifically, the framework jointly quantifies array-domain spatial non-stationarity (SnS), and spatial degrees of freedom (DoF) in line-of-sight, 3GPP non-line-of-sight, and isotropic-scattering propagation environments. 
For SnS, a novel Power-balanced, Visibility-aware Correlation-Matrix Distance (PoVi-CMD) and a two-stage subarray-screening procedure are introduced. 
For DoF, the R\'enyi-2 effective rank is adopted, and tractable spatial-correlation expressions under isotropic scattering are developed for efficient DoF analysis. 
Furthermore, a realizable antenna port mode is introduced to connect SnS with DoF. 
Numerical results reveal that curvature and propagation support are the primary determinants of both SnS and DoF in HoloCuRA: array domain SnS determines whether subarray statistics can be treated as locally consistent, whereas DoF limits the global spatial modes.
The findings provide useful guidance for low-altitude antenna-system design.
\end{abstract}


\section{Introduction}
\IEEEPARstart{E}{lectrically} large, densely sampled antenna array apertures provide a powerful means to control radiation, focusing, and spatial selectivity by approaching the continuous-aperture regime. 
Holographic multiple-input multiple-output (HMIMO) is a representative realization of this paradigm, offering high aperture efficiency together with substantial geometric flexibility~\cite{Ref_huang2020holographic,boyadi2022reconfigurable}. 
Nevertheless, a perfectly planar aperture is rarely guaranteed in realistic deployments, since bending and warping arise from thin substrates, tiled metasurfaces, tolerances, and thermo-mechanical stress, which motivates curvature reconfigurable aperture~\cite{Ref_ConformalRadar}.
Recent advances in flexible intelligent metasurfaces (FIMs), also known as morphable intelligent metasurfaces, provide a natural hardware pathway toward curvature-reconfigurable apertures~\cite{an2025flexible,Yoo2022ConformalMTS}.
In this work, we specifically consider continuous, curvature-controllable, reconfigurable aperture. Accordingly, such curvature fundamentally modifies the array manifold, local visibility, and propagation support~\cite{Ref_ConformalRadar}. 
Consequently, planar, far-field, and quasi-stationary assumptions can be violated in Holographic Curvature-Reconfigurable Aperture (HoloCuRA), making classical conclusions inadequate. It is therefore essential to systematically quantify the resulting spatial propagation characteristics. 


\subsection{Prior Works}
Before analyzing these characteristics in detail, we briefly review prior work on spatial non-stationarity (SnS), one of the channel properties most directly affected by aperture curvature.
In the literature, SnS is used in two closely related but not identical senses. From a classical stochastic signal-processing perspective, it refers to the violation of wide-sense stationarity (WSS), i.e., the loss of translation invariance in a chosen domain such as time, frequency, or space~\cite{Matz2005NonWSSUS,Gao2013AsilomarMeasModels}.
By contrast, in propagation and channel-measurement studies, SnS is typically reflected by aperture-dependent variations in second-order statistics, such as received power and cluster visibility~\cite{Carvalho2020NonStationarities}.
For HoloCuRA, these two viewpoints naturally coincide, since the electrically large aperture and curvature-induced geometry variations can invalidate spatial stationarity even within the array domain.

Existing SnS studies can be broadly grouped into three categories:
(i) measurement-driven characterization across time, frequency, space, and deployment locations, often accompanied by SnS-aware channel modeling; 
(ii) system-level SnS-aware design, for extra-large MIMO (XL-MIMO) that explicitly incorporates SnS into transceiver design and resource management; and
(iii) metric and criterion development, for which a unified and widely accepted standard remains unavailable. 
For the first category, extensive measurement campaigns have reported pronounced SnS in XL-MIMO and distributed-MIMO channels~\cite{Yuan2023SnSNF,3DSnSBian,Xu2024ICCWs_THZ132GHz,Gao2015DoAllAntennas,Xu2025EuCAPNonWSS}.
To model the underlying mechanism that some multipath components (MPCs) are observable only over portions of the aperture, the concept of visibility region (VR) was introduced and has since been widely adopted in SnS-aware channel models~\cite{Flordelis2020COST2100}.
Building on this insight, research in the second category has primarily explored the new opportunities brought by SnS, especially VR~\cite{Li2015CapacityNonWSS}. 
Rather than enforcing a global stationarity assumption, these studies exploit the fact that different subarrays can observe distinct dominant paths and VRs, which naturally leads to subarray-wise processing and structure-aware channel state information (CSI) acquisition~\cite{Pang2025TFSExtrapolation,Chen2024NonStationaryCE_XL,Wang2025TCCN_FDVRWideband}.
The third category of research, which is a main focus of this paper, is to quantify SnS through practical and computable metrics.
A representative approach is to measure how the channel’s spatial structure evolves by computing distances between correlation matrices estimated at different times/positions, leading to the widely used correlation matrix distance (CMD)~\cite{CMD1,CMD2}. 
However, subsequent analyses of CMD have revealed that it may underestimate non-stationarity for full-rank covariance matrices. 
Accordingly, eigenvalue-normalized CMD has been proposed to mitigate these limitations~\cite{NCMD}.
Taken together, these studies show that, although existing SnS metrics are operationally useful, their characterization capability remains incomplete, which motivates the metric design and analysis developed in this paper.

Another key spatial descriptor is the DoF, which quantifies the number of independent spatial modes supported by a given aperture--propagation configuration and thus determines the effective spatial rank. 
Although DoF scaling laws are well understood for canonical apertures under far-field propagation, including linear, planar, volumetric, and spherical geometries~\cite{Ref_SunTao2022,Ref_PizzoMarzetta2020,Ref_poon2005degrees}, these results do not directly address curvature-reconfigurable holographic apertures. Meanwhile, most existing holographic-aperture analyses still rely on planar abstractions~\cite{Ref_PizzoMarzetta2020}.
Such results do not directly extend to conformal apertures or HoloCuRA, where curvature changes local orientation, visibility, and the effective aperture presented to the propagation field, thus requiring geometry-aware modeling~\cite{Ref_ConformalRadar,Pelham2017PredictingGain,Yoo2022ConformalMetasurface}. 
While conformal-antenna studies have extensively characterized radiation properties such as beam patterns and efficiency~\cite{Ref_ConformalRadar,allard2003radiation,cheng2013millimeter}, research on flexible and conformal antennas/arrays has still been largely radiator-oriented, focusing on deformation-aware array operation, conformal antenna realization, and curved metasurface-array design rather than channel-level spatial characterization~\cite{tang2020flexible,Braaten2013SELFLEX,Sayem2019TransparentConformal,Yoo2022ConformalMTS}.
In contrast, a systematic characterization of how curvature affects both SnS and DoF in HoloCuRA remains largely open.

This gap is fundamental because curvature reshapes HoloCuRA channels at both the local-statistical and global-modal levels. Accordingly, SnS and DoF are adopted here as two complementary descriptors of spatial behavior: SnS and DoF capture two complementary spatial aspects of HoloCuRA channels: SnS reflects the local consistency of second-order statistics along the aperture, whereas DoF characterizes the number of globally resolvable spatial modes. Since ignoring SnS can bias subarray modeling and beam management, while ignoring DoF can overestimate spatial parallelism, both must be treated jointly. In HoloCuRA, they are further coupled by curvature through the underlying three-dimensional (3D) propagation geometry. 

To expose this coupling, we consider the representative low-altitude scenario illustrated in Fig.~\ref{fig1}. Here, HoloCuRA is deployed on terrestrial or aerial base stations with curved or space-constrained surfaces.  Wireless links with unmanned aerial vehicles (UAVs), aircraft, and nearby ground or maritime users encounter rapidly varying elevation and azimuth angles, intermittent blockage, and mixed propagation with dominant line-of-sight (LoS) and local scattering~\cite{A2Gchannelmodel,Motlagh2016UAVIoT,Xu2025LLMEmpoweredLAE,Sun2026MaritimeTWC}. Such conditions accentuate curvature-induced spatial variations and make low-altitude HoloCuRA a particularly relevant setting for spatial characterization.

\subsection{Our Contributions}

Despite substantial prior work on non-stationary channels and conformal antennas, a rigorous propagation-oriented characterization of curvature-reconfigurable holographic apertures remains lacking. Existing SnS studies mainly address time- or user-position-induced variations, with limited attention to array-domain SnS on curved dense apertures. Likewise, most available DoF analyses are still rooted in conventional linear or planar apertures. Meanwhile, related conformal-antenna studies remain largely radiator-oriented, focusing primarily on matching and radiation performance rather than channel-level spatial behavior. As a result, a unified understanding of how curvature reshapes visibility, spatial correlation, array-domain SnS, and effective spatial DoF in HoloCuRA is still missing. To address this gap, we develop a visibility-aware spatial characterization framework for HoloCuRA. The main contributions are summarized as follows.

{%
\makeatletter
\long\def\@makecaption#1#2{%
  \vskip\abovecaptionskip
  \sbox\@tempboxa{\footnotesize #1. #2}%
  \ifdim \wd\@tempboxa >\hsize
    \footnotesize #1. #2\par
  \else
    \global\@minipagefalse
    \hb@xt@\hsize{\hfil\box\@tempboxa\hfil}%
  \fi
  \vskip\belowcaptionskip}
\makeatother

\begin{figure}[t!]
    \centering
    \includegraphics[width=0.49\textwidth]{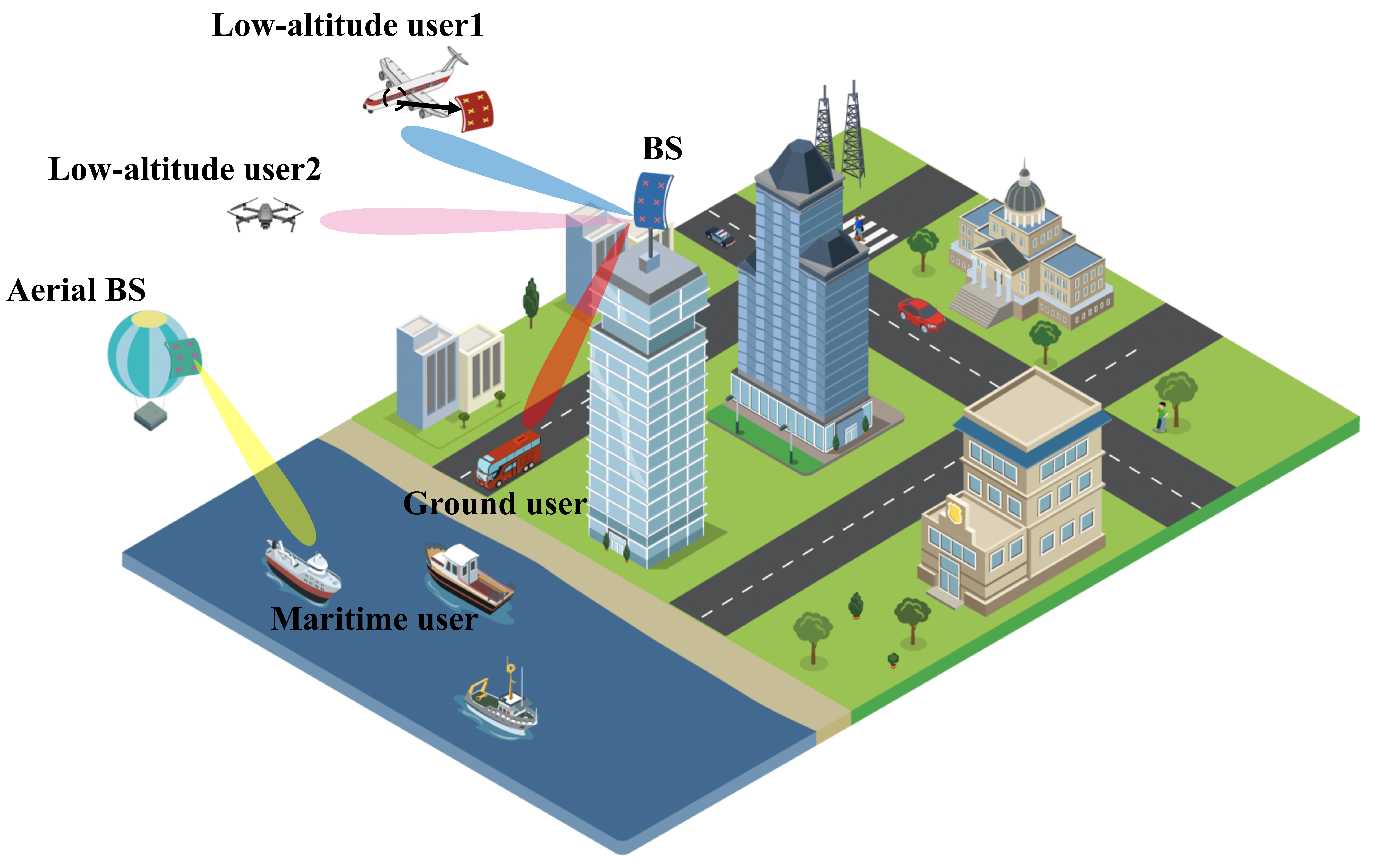}
    \caption{Illustration of a low-altitude HoloCuRA system.}
    \label{fig1}
\end{figure}
}

\begin{itemize}

\item We establish a unified spatial characterization framework for HoloCuRA that jointly analyzes array-domain SnS and DoF across three canonical propagation environments: LoS, 3GPP NLoS, and isotropic scattering. The framework connects local statistical admissibility to full-aperture spatial behavior under a common curvature-aware setting.

\item We propose the Power-balanced and Visibility-aware Correlation Matrix Distance (PoVi-CMD, $d_{\mathrm{PoVi}}$) for curved apertures. Unlike conventional CMD-type measures, PoVi-CMD captures both correlation-structure variation and visibility-induced power imbalance. Based on this metric, we further develop a two-stage SnS procedure consisting of local subarray stationarity screening and full-aperture SnS mapping.

\item We derive tractable spatial-correlation kernels and correlation-matrix expressions for 1D/2D HoloCuRA, with particular emphasis on isotropic scattering, where an efficient analytical DoF characterization is obtained through a closed-form baseline with a bounded one-dimensional correction. This places the DoF analysis on an explicit propagation-correlation foundation and, in the isotropic-scattering case, makes the impact of curvature and propagation support analytically transparent.

\end{itemize}

In this work, the holographic nature of HoloCuRA is essential rather than incidental. Dense holographic sampling pushes the aperture toward the continuous-aperture regime, suppresses discretization artifacts, and makes the resulting SnS and DoF behavior primarily geometry- and propagation-driven. At the same time, it also provides a practical discrete approximation to the continuous-aperture correlation model and a natural bridge to port limited implementations.


\subsection{Organization and Notation}
\textit{Organization}: The remainder of this paper is organized as follows. Section II introduces the HoloCuRA geometries. Section III presents the adopted SnS/DoF metrics and the proposed PoVi-CMD. Section IV studies LoS spatial characteristics, including VR determination, local stationarity screening, and array-domain SnS mapping. Section V considers 3GPP CDL-A channels. Section VI derives half-space isotropic-scattering correlation expressions and discusses their SnS/DoF implications. Section VII bridges SnS and DoF via realizable port modes. Section VIII concludes the paper.

\textit{Notation}: Bold lowercase/uppercase letters denote vectors/matrices; $(\cdot)^T$ and $(\cdot)^H$ denote transpose and conjugate transpose; $\mathbb{E}\{\cdot\}$ denotes expectation; $\mathrm{tr}(\cdot)$ and $\|\cdot\|_F$ denote trace and Frobenius norm; $\odot$ denotes the Hadamard product; and $a \mid b$ means that $a$ divides $b$.


\section{System Model}


We consider a BS equipped with HoloCuRA that serves single-antenna users in 3D space. The discrete coordinates below represent a uniformly sampled realization of HoloCuRA. This section defines the one dimension (1D)/ two dimension (2D) aperture geometry and the associated exact and far-field array responses used in the subsequent SnS and DoF analysis. Unless otherwise stated, the LoS analysis relies on the exact spherical-wave distance, whereas the far-field approximation is invoked later when appropriate for the NLoS and isotropic-scattering settings.

\subsection{1D HoloCuRA Geometry}

Consider a 1D HoloCuRA with total length $L$ and curvature radius $R$, vertically deployed in the YZ-plane as shown in Fig.~\ref{fig2}. Let $N$ denote the number of uniformly spaced samples along the curved aperture, with arc spacing $d_{yz}$. A single-antenna user is located at
\begin{equation}
\mathbf u
=
\bigl[
r\sin\theta\cos\phi,\;
r\sin\theta\sin\phi,\;
r\cos\theta
\bigr],
\end{equation}
where $r$ denotes the user-to-origin distance, $\theta$ is the zenith angle, and $\phi$ is the azimuth angle. The position of the $n$-th element is
\begin{equation}
\mathbf p_n
=
\bigl[
0,\;
R\cos(\beta-\alpha_n)-R\cos\beta,\;
R\sin(\beta-\alpha_n)
\bigr],
\label{eq:p_n_1d}
\end{equation}
where
\begin{equation}
\beta=\frac{L}{2R}\in\Bigl[0,\frac{\pi}{2}\Bigr],
\qquad
\alpha_n=\frac{(n-1)L}{(N-1)R},\quad n=1,\ldots,N.
\label{eq:alpha_n}
\end{equation}

The exact user-to-element distance is
\begin{equation}
r_{u,n}=\|\mathbf u-\mathbf p_n\|,
\label{eq:run_exact}
\end{equation}
\begin{equation}
r_{u,n} = \sqrt{
\begin{aligned}
&\bigl[r\,\sin(\theta)\cos(\phi)\bigr]^2
\\+&\Bigl[r\,\sin(\theta)\sin(\phi) - R\,\cos(\beta-\alpha_n) 
    + R\,\cos(\beta)\Bigr]^2
 \\+&\bigl[r\,\cos(\theta) 
         - R\,\sin(\beta-\alpha_n)\bigr]^2. 
         \end{aligned}
         } 
\end{equation}

Accordingly, the exact LoS array response, referenced to the array origin can be expressed as
\begin{equation}
\mathbf a_{\rm}(r,\theta,\phi)
=
\frac{1}{\sqrt N}
\bigl[
e^{-j\frac{2\pi}{\lambda}(r_{u,1}-r)},
\ldots,
e^{-j\frac{2\pi}{\lambda}(r_{u,N}-r)}
\bigr]^{\mathrm T},
\label{eq:a1d_exact}
\end{equation}
where $\lambda$ is the carrier wavelength.
Under the classical far-field condition $r\gg L$, a first-order expansion of \eqref{eq:run_exact} yields
\begin{equation}
r_{u,n}\approx r-\Delta_n,
\label{eq:run_ff}
\end{equation}
with
\begin{equation}
\resizebox{0.96\columnwidth}{!}{$
\Delta_n
=
R\!\left[
\sin\theta\sin\phi\bigl(\cos(\beta-\alpha_n)-\cos\beta\bigr)
+
\cos\theta\sin(\beta-\alpha_n)
\right].
$}
\label{eq:Delta_n}
\end{equation}

The corresponding far-field array manifold is
\begin{equation}
\mathbf a(\theta,\phi)
=
\frac{1}{\sqrt N}
\bigl[
e^{j\frac{2\pi}{\lambda}\Delta_1},
\ldots,
e^{j\frac{2\pi}{\lambda}\Delta_N}
\bigr]^{\mathrm T}.
\label{eq:arcULA_steer2}
\end{equation}

{%
\makeatletter
\long\def\@makecaption#1#2{%
  \vskip\abovecaptionskip
  \sbox\@tempboxa{\footnotesize #1. #2}%
  \ifdim \wd\@tempboxa >\hsize
    \footnotesize #1. #2\par
  \else
    \global\@minipagefalse
    \hb@xt@\hsize{\hfil\box\@tempboxa\hfil}%
  \fi
  \vskip\belowcaptionskip}
\makeatother

\begin{figure}[t!]
    \centering
    \includegraphics[width=0.49\textwidth]{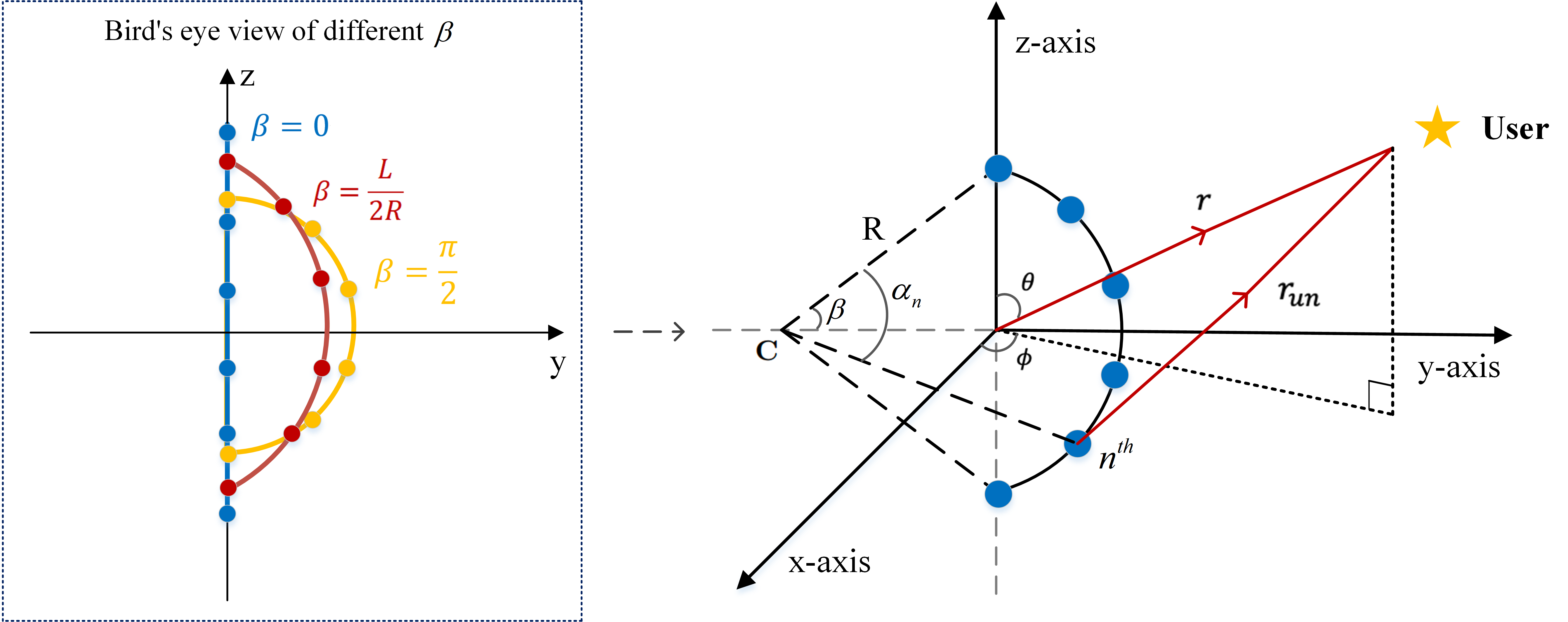}
    \caption{Geometry and top view of 1D HoloCuRA in 3D space.}
    \label{fig2}
\end{figure}
}

\subsection{2D HoloCuRA Geometry}

\begin{figure}[t!]
	\begin{centering}
		\includegraphics[width=0.49\textwidth]{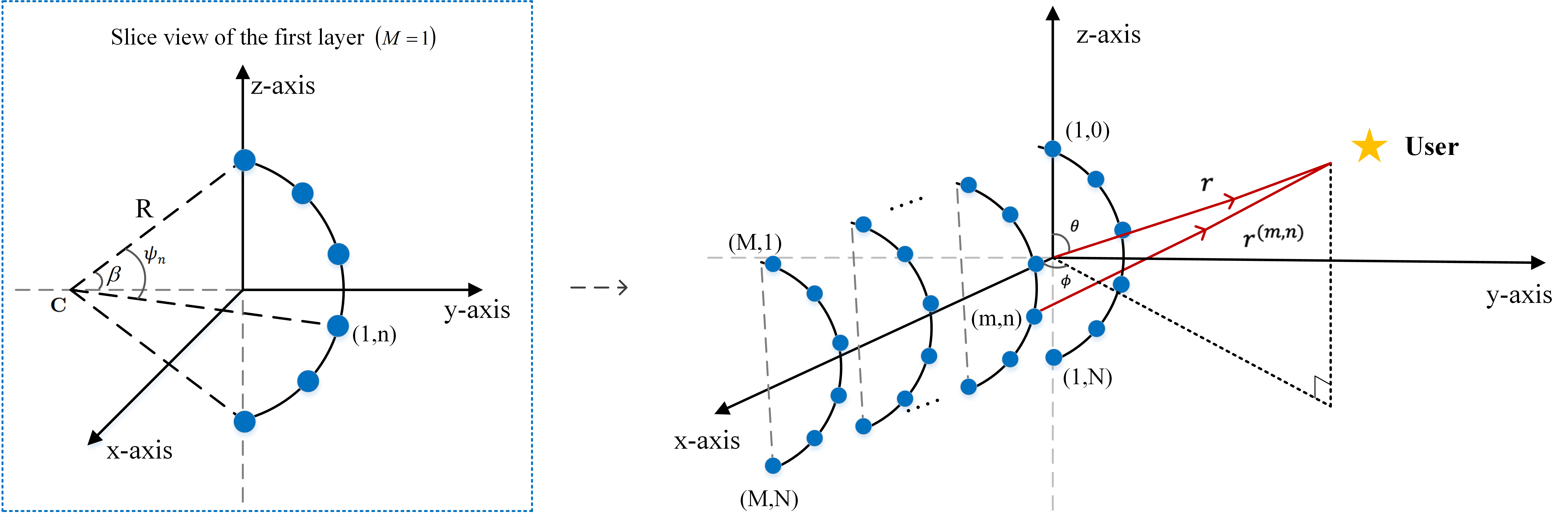}
        \captionsetup{font=footnotesize, labelsep=period}
		\caption{Geometry and the first-layer slice view of 2D HoloCuRA in 3D space.}\label{fig3}
	\end{centering}
\end{figure}

In real-word deployments, planar arrays are more common than linear arrays. Accordingly, 1D HoloCuRA can be extended to 2D HoloCuRA to improve coverage, as illustrated in Fig.~\ref{fig3}.
Specifically, the 2D aperture consists of $M$ identical 1D HoloCuRA segments placed at $x=md_x$, $m=1,\ldots,M$, with uniform inter-segment spacing $d_x$. For practical relevance and uniform aperture sampling, we assume $d_x=d_{yz}$ throughout.

The $(m,n)$-th element is located at
\begin{equation}
\mathbf p_{m,n}
=
\bigl(
md_x,\;
R\cos(\beta-\psi_n)-R\cos\beta,\;
R\sin(\beta-\psi_n)
\bigr),
\label{eq:element_coord}
\end{equation}
where $\psi_n =\frac{(n-1)\,L}{(N-1)\,R}\in[0,\pi]$ parameterizes the $n$-th element on each segment. The symbol $\psi_n$ is retained to match Fig.~\ref{fig3}; it plays the same geometric role in 2D as $\alpha_n$ in the 1D case. The exact user-to-element distance is
\begin{equation}
r_u^{(m,n)}=\|\mathbf u-\mathbf p_{m,n}\|.
\label{eq:rumne_exact}
\end{equation}
\begin{equation}
r_u^{(m,n)} = \sqrt{
    \begin{aligned}
    &[r\sin(\theta)\cos(\phi) - md_x]^2 \\
    +&\left[r\sin(\theta)\sin(\phi) - R\cos(\beta-\psi_n) + R\cos(\beta)\right]^2 \\
    +&\left[r\cos(\theta) - R\sin(\beta-\psi_n)\right]^2.
    \end{aligned}
}
\end{equation}

 The exact 2D LoS array response, again referenced to the array origin can be written by
\begin{equation}
\resizebox{0.96\columnwidth}{!}{$
\mathbf a_{\mathrm{2D}}(r,\theta,\phi)
=
\frac{1}{\sqrt{MN}}
\bigl[
e^{-j\frac{2\pi}{\lambda}(r_u^{(1,1)}-r)},
\ldots,
e^{-j\frac{2\pi}{\lambda}(r_u^{(M,N)}-r)}
\bigr]^{\mathrm T}
$}.
\label{eq:a2d_exact}
\end{equation}

When the user is in the far field relative to both aperture dimensions, i.e.,
$r\gg \max\{L,Md_x\}$. A first-order expansion gives
\begin{equation}
r_u^{(m,n)}\approx r-\Delta_{m,n},
\label{eq:rumne_ff}
\end{equation}
where
\begin{equation}
\begin{aligned}
  \Delta_{m,n} &= m\,d_x\,\sin(\theta)\,\cos(\phi) 
  \;\\&+\; R\Bigl[\sin(\theta)\,\sin(\phi)\,\bigl(\cos(\beta-\psi_n) - \cos(\beta)\bigr) 
    \;\\&+\; \cos(\theta)\,\sin(\beta-\psi_n)\Bigr].
\end{aligned}
\label{eq:Delta_mn}
\end{equation}

The corresponding far-field array manifold is
\begin{equation}
\mathbf a_{\mathrm{2D}}(\theta,\phi)
=
\frac{1}{\sqrt{MN}}
\bigl[
e^{j\frac{2\pi}{\lambda}\Delta_{1,1}},
\ldots,
e^{j\frac{2\pi}{\lambda}\Delta_{M,N}}
\bigr]^{\mathrm T}.
\label{eq:steer_ura_compact}
\end{equation}

The above 1D/2D geometries define the aperture manifolds used in the subsequent SnS and DoF analysis. In particular, curvature enters through the element coordinates and thereby affects visibility, spatial correlation, and propagation support across the aperture.

\section{Spatial Characterization Metrics}

Building on the aperture manifolds in Section II, we characterize HoloCuRA through two complementary descriptors: array-domain SnS and effective spatial DoF. SnS is quantified by the proposed PoVi-CMD, whereas DoF is measured by the R\'enyi-2 effective rank of the full-aperture correlation matrix. 


\subsection{PoVi-CMD for VR-Aware SnS}

Classical CMD measures the similarity between correlation structures, but it normalizes away absolute power and therefore cannot reflect VR-induced power inconsistency across subarrays. For HoloCuRA, however, visibility interruption alters not only the correlation structure but also the power supported by each visible subspace. Motivated by this, we introduce a Power-Balanced and Visibility-Aware CMD, which jointly captures structural decorrelation and visibility-induced power mismatch.

Consider either a 1D HoloCuRA with $N_{\mathrm{ant}}=N$ elements or a 2D HoloCuRA with $N_{\mathrm{ant}}=MN$ elements. Let
$\mathbf h=[h_1,\ldots,h_{N_{\mathrm{ant}}}]^{\mathrm T}\in\mathbb C^{N_{\mathrm{ant}}\times 1}$ denote the full-array channel, and partition the aperture into $K$ equal-size subarrays $\{S_k\}_{k=1}^{K}$, where
$S_k=\{i_{k,1},\ldots,i_{k,N_{\mathrm{sub}}}\}$ and $N_{\mathrm{sub}}=|S_k|$. The corresponding subarray channel is
$\mathbf h_k=[h_{i_{k,1}},\ldots,h_{i_{k,N_{\mathrm{sub}}}}]^{\mathrm T}$.
For a scenario-dependent visible set $\mathcal V_p\subseteq\{1,\ldots,N_{\mathrm{ant}}\}$, define the binary mask
$\mathbf v_k=[v_{k,1},\ldots,v_{k,N_{\mathrm{sub}}}]^{\mathrm T}$ with $v_{k,j}=1$ if $i_{k,j}\in\mathcal V_p$ and $v_{k,j}=0$ otherwise. The VR-gated (visible) channel is then
\begin{equation}
\mathbf h_k^{\mathrm{vr}}=\mathbf h_k\odot \mathbf v_k
\label{eq:vis-channel}
\end{equation}
which yields the visible spatial correlations
\begin{equation}
\mathbf R_{kk}^{\mathrm{vr}}=\mathbb E\{\mathbf h_k^{\mathrm{vr}}(\mathbf h_k^{\mathrm{vr}})^{\!H}\},
\qquad
\mathbf R_{k\ell}^{\mathrm{vr}}=\mathbb E\{\mathbf h_k^{\mathrm{vr}}(\mathbf h_\ell^{\mathrm{vr}})^{\!H}\}.
\label{eq:vis-covs}
\end{equation}
where $\mathbf{R}_{kk}^{\mathrm{vr}}$ is the auto-correlation matrix of the VR channel vector $\mathbf{h}_k^{\mathrm{vr}}$,
and $\mathbf{R}_{k\ell}^{\mathrm{vr}}$ is the corresponding cross-correlation matrix between $\mathbf{h}_k^{\mathrm{vr}}$ and $\mathbf{h}_\ell^{\mathrm{vr}}$, $k$ and $\ell$ index two (possibly different) channel vectors (e.g., the $k$-th and $\ell$-th subarrays/links).
Then, the defined PoVi-CMD with a power exponent $q\ge1$ can be formulated as
\begin{equation}
\gamma_{k\ell}
=
\frac{\operatorname{tr}(\mathbf R_{kk}^{\mathrm{vr}}\mathbf R_{\ell\ell}^{\mathrm{vr}})}
{\|\mathbf R_{kk}^{\mathrm{vr}}\|_F\,\|\mathbf R_{\ell\ell}^{\mathrm{vr}}\|_F},
\label{eq:gamma}
\end{equation}
\begin{equation}
\alpha_{k\ell}
=
\frac{\|\mathbf R_{kk}^{\mathrm{vr}}\|_F}{\|\mathbf R_{\ell\ell}^{\mathrm{vr}}\|_F},
\label{eq:alpha}
\end{equation}
\begin{equation}
f_q(\alpha)=\frac{2}{\alpha^q+\alpha^{-q}}
=
\operatorname{sech}\!\big(q|\ln\alpha|\big),\qquad q\ge1,
\label{eq:fp}
\end{equation}
and
\begin{equation}
d_{\mathrm{PoVi}}^{k\ell}(q)
=
1-\gamma_{k\ell}f_q(\alpha_{k\ell})\in[0,1].
\label{eq:dPoVi}
\end{equation}

Here, $\gamma_{k\ell}$ measures correlation-structure coherence, $\alpha_{k\ell}$ captures the visible-power ratio, and $f_q(\alpha_{k\ell})$ penalizes log-power mismatch, attaining $1$ at $\alpha_{k\ell}=1$ and vanishing as $\alpha_{k\ell}\to0$ or $\alpha_{k\ell}\to\infty$. Unless otherwise stated, $d_{\mathrm{PoVi}}^{k\ell}$ denotes $d_{\mathrm{PoVi}}^{k\ell}(q)$ for a fixed design choice of $q$.

\subsection{Effective Spatial DoF}

To quantify the global modal support of HoloCuRA, we adopt the R\'enyi-2 effective rank of the full-aperture spatial correlation matrix
$\mathbf R=\mathbb E\{\mathbf h\mathbf h^H\}$. Let $\{\lambda_i\}$ denote the eigenvalues of $\mathbf R$, and define
$p_i=\lambda_i/\sum_j\lambda_j$. The R\'enyi-2 entropy is
$H_2=-\log\!\left(\sum_i p_i^2\right)$, which gives the effective rank~\cite{Renyi1961Entropy}
\begin{equation}
\exp(H_2)
=
\frac{1}{\sum_i p_i^2}
=
\frac{\left(\sum_i \lambda_i\right)^2}{\sum_i \lambda_i^2}
=
\frac{\big(\mathrm{tr}(\mathbf R)\big)^2}{\|\mathbf R\|_F^2}.
\label{eq:renyi2_eff_rank}
\end{equation}

Since \eqref{eq:renyi2_eff_rank} is invariant to any positive scaling of $\mathbf R$, it measures eigenvalue spread rather than absolute power and thus quantifies the effective number of supported spatial modes. Unless otherwise stated, DoF is computed from the full $\mathbf R$ without VR gating, since the objective here is to characterize the physical modal ceiling of the aperture--propagation pair; the associated visibility-limited realizability loss is treated later in Section VII. We also report the normalized DoF
\begin{equation}
n_{\mathrm{DoF}}=\frac{N_{\mathrm{DoF}}}{N_{\mathrm{ant}}}.
\end{equation}

In summary, $d_{\mathrm{PoVi}}$ is stationarity-oriented, since it measures cross-subarray statistical consistency under VR gating, whereas $N_{\mathrm{DoF}}$ is mode-oriented, since it summarizes the total number of spatial modes supported by the full aperture. Together, they provide a compact and physically grounded characterization framework for the subsequent LoS, NLoS, and isotropic-scattering analysis.

\section{LoS Spatial Characterization: Visibility, SnS, and DoF}

With the metrics in Section III in place, we turn to the LoS case, where the impact of curvature is most directly governed by aperture geometry, spherical-wave propagation, and self-occlusion. The analysis proceeds in two stages. We first derive an explicit visible-region (VR) criterion and use it to determine admissible subarray partitions through a local stationarity test. We then characterize the resulting array-domain SnS over the full aperture and complement it with a DoF view under LoS-dominant sparse propagation.

\subsection{Geometric Visible-Region Determination}

\subsubsection{VR Criterion for 1D HoloCuRA}
\label{subsec:VR_ULA}

For 1D HoloCuRA in the $y$--$z$ plane, LoS visibility is governed by convex self-occlusion. Let the $n$-th element be located at $\mathbf p_n$ and let the effective center of curvature be $\mathbf C=(0,-R\cos\beta,0)$. The outward unit normal at element $n$ is
\begin{equation}
\hat{\mathbf n}_n=\frac{\mathbf p_n-\mathbf C}{R}
=\big[0,\cos\gamma_n,\sin\gamma_n\big], \quad
\gamma_n=\beta-\alpha_n .
\label{eq:nn}
\end{equation}

Let the user position be $\mathbf u$. By the tangent-plane test, element $n$ is visible if
\begin{equation}
(\mathbf u-\mathbf p_n)\!\cdot\!\hat{\mathbf n}_n\ge 0
\ \Leftrightarrow\
(\mathbf u-\mathbf C)\!\cdot\!\hat{\mathbf n}_n\ge R .
\label{eq:tangent}
\end{equation}

Substituting \eqref{eq:nn} into \eqref{eq:tangent} yields
\begin{equation}
r\,g_n(\theta,\phi)\ge R\bigl(1-\cos\beta\cos\gamma_n\bigr),
\label{eq:ineq-raw}
\end{equation}
where
\begin{equation}
g_n(\theta,\phi)=
\sin\theta\sin\phi\cos\gamma_n+\cos\theta\sin\gamma_n,
\label{eq:g-Gamma}
\end{equation}
\begin{equation}
\Gamma_n(r,\beta,\gamma_n)=\frac{R}{r}\bigl(1-\cos\beta\cos\gamma_n\bigr).
\label{eq:g-gamma}
\end{equation}

Hence, the VR condition is
\begin{equation}
g_n(\theta,\phi)\ge \Gamma_n(r,\beta,\gamma_n).
\label{eq:ULA-VR}
\end{equation}

Under free-space LoS propagation,
\[
h_n^{\mathrm{LoS}}
=
\sqrt{G_tG_r}\,\frac{\lambda}{4\pi r_{un}}
e^{-j\frac{2\pi}{\lambda}r_{un}},
\qquad
r_{un}=\|\mathbf u-\mathbf p_n\|,
\]
where $G_t$ and $G_r$ are the transmit and receive antenna gains and are set to unity. For the adopted 1D deployment, the binary VR mask $v_n^{\mathrm{LoS}}\in\{0,1\}$ is activated at $\phi=\pi/2$ through \eqref{eq:ULA-VR}, whereas for $\phi\neq\pi/2$ we set $v_n^{\mathrm{LoS}}=1$. The VR-aware LoS channel is then
\begin{equation}
\mathbf h^{\mathrm{LoS\mbox{-}VR}}
=
\mathbf v^{\mathrm{LoS}}\odot\mathbf h^{\mathrm{LoS}} .
\label{eq:hVR-RVR}
\end{equation}

\subsubsection{Extension to 2D HoloCuRA}
\label{subsec:VR_URA}

For 2D HoloCuRA, the same geometric criterion applies row-wise along the curved dimension, but the visibility test is evaluated over the full azimuth range. In implementation, we compute the mask along the arc direction and replicate it across the $x$-dimension. The resulting mask directly gates the LoS channel, yielding the VR-aware 2D response without further geometric modification.

\subsubsection{Geometric Implications of the VR Criterion}

The criterion in \eqref{eq:ULA-VR} highlights two vital properties. First, there is no single distance $r$ that guarantees full visibility for all look directions, because visibility depends jointly on $(\theta,\phi)$, $r$, and $\beta$. Second, even in the far-field limit ($r\to\infty$), full visibility is not automatic; the outward half-space condition still requires $g_n(\theta,\phi)\ge 0$. In the planar limit $\beta\to 0$, \eqref{eq:ULA-VR} reduces to the familiar half-space test.

\subsection{Local SnS Criterion and Admissible Subarray Design}

\subsubsection{Local Stationarity Criterion}

In LoS, VR-induced non-stationarity first appears locally within each subarray. We therefore determine admissible subarray partitions before evaluating full-aperture SnS. For 1D HoloCuRA, the aperture is divided into $K$ contiguous subarrays of equal size
\begin{equation}
N_{\mathrm{sub}}=\frac{N}{K},
\qquad K\mid N .
\end{equation}

For 2D HoloCuRA, a grid partition is specified by $(g_z,g_x)$ such that
\begin{equation}
K=g_zg_x,\qquad g_z\mid N,\qquad g_x\mid M,
\end{equation}
with tile size
\begin{equation}
\mathrm{tileZ}=\frac{N}{g_z},\qquad
\mathrm{tileX}=\frac{M}{g_x},\qquad
N_{\mathrm{sub}}=\mathrm{tileZ}\cdot\mathrm{tileX}.
\end{equation}

For a fixed $K$, we consider all feasible factor pairs $(g_z,g_x)$ satisfying the above divisibility
constraints. Thus, larger $g_z$ produces thinner strips along $z$ (smaller $\mathrm{tileZ}$), while larger $g_x$ produces
narrower tiles along $x$ (smaller $\mathrm{tileX}$). Each tile (subarray) contains $N_{\mathrm{sub}}=\mathrm{tileZ}\cdot \mathrm{tileX}$ elements.
\footnote{For example, with $K=64$ and $N_z=N_x=128$, ``Z-cut'' corresponds to $\mathrm{Grid}\ 64\times 1$
(tiles of size $2\times 128$), whereas ``X-cut'' corresponds to $\mathrm{Grid}\ 1\times 64$ (tiles of size
$128\times 2$). Other grids (e.g., $2\times 32$, $8\times 8$) cut both dimensions and yield $\mathrm{tileZ}\times \mathrm{tileX}$ tiles.}

Let $\Omega=(\theta,\phi)$ denote a sampled half-space direction, and let $h_i(\Omega;\beta,r)$ be the VR-aware LoS channel at element $i$, with invisible elements set to zero. Define the element-wise power
\begin{equation}
a_i(\Omega)\triangleq |h_i(\Omega)|^2 .
\end{equation}

For subarray $S_k$, let $\mu_k(\Omega)$ and $v_k(\Omega)$ be the mean and variance of $\{a_i(\Omega):i\in S_k\}$, and define the relative variance
\begin{equation}
\eta_k(\Omega)\triangleq \frac{v_k(\Omega)}{\mu_k(\Omega)^2},
\qquad \mu_k(\Omega)>0.
\end{equation}

Direction $\Omega$ is declared locally passing if $\eta_k(\Omega)\le 1/e$.\footnote{The $1/e$ threshold is commonly used
in channel modeling~\cite{Sun2026MaritimeTWC} and provides a practical boundary between locally flat and
visibly non-uniform power profiles.} Over all sampled directions $\{\Omega_t\}_{t=1}^{N_\Omega}$, define the active set
\begin{equation}
V_k\triangleq \{t:\mu_k(\Omega_t)>0\}
\end{equation}
and the pass probability
\begin{equation}
P_k\triangleq
\frac{1}{|V_k|}
\sum_{t\in V_k}
\mathbf 1\!\left\{\eta_k(\Omega_t)\le \frac{1}{e}\right\}.
\end{equation}

Subarray $k$ is declared locally stationary if $P_k\ge 0.7$.\footnote{The threshold $0.7$ is adopted as a strong-majority rule: it requires that the local flatness criterion holds for most ($\geq 70\%$) of the visible directions, while a small  ($\leq 30\%$) fraction of outliers is tolerated.} This two-level test first checks local flatness for each direction and then retains only partitions that remain stable over a strong majority of visible directions.

\subsubsection{Partition Candidates and Evaluation Setup}

Low-altitude links in the millimeter-wave/sub-THz regime are often LoS-dominant or NLoS with directionally sparse
scattering. The small wavelength also makes a fixed physical aperture electrically large
($D/\lambda$), which pushes the radiating near-field (Fresnel region) to meter-/tens-of-meters
link ranges, thereby making array-domain SnS easier to excite and observe. In
addition, the selected ranges should be consistent with classical aperture distance bounds.
Building on this, we evaluate the local criterion at $f_c=30~\mathrm{GHz}$ with fixed aperture length $L=0.32~\mathrm{m}$ and curvature $\beta\in(0,\pi/2]$. Following standard aperture-distance criteria, we use the chord length as the effective aperture size, whose extrema are
\begin{equation}
D_{\max}=L,\qquad
D_{\min}=\frac{2L}{\pi}\approx 0.2037~\mathrm{m}.
\label{eq:Dextrema}
\end{equation}

The reactive near-field bound and Rayleigh distance~\cite{Ref_sun2025differentiate} are
\begin{equation}
R_{\mathrm{react}}(D)\approx 0.62\sqrt{\frac{D^3}{\lambda}},
\qquad
R_{\mathrm{Rayleigh}}(D)=\frac{2D^2}{\lambda}.
\label{eq:bounds}
\end{equation}

Taking conservative extrema over $\beta$ gives
\begin{equation}
R_{\mathrm{react,max}}
=
R_{\mathrm{react}}(D_{\max})
\approx 1.12~\mathrm{m},
\label{eq:extrema_nf_ff1}
\end{equation}
\begin{equation}
R_{\mathrm{Rayleigh,min}}
=
R_{\mathrm{Rayleigh}}(D_{\min})
\approx 8.30~\mathrm{m}.
\label{eq:extrema_nf_ff}
\end{equation}

Accordingly, we use $r\in\{2,8,100\}~\mathrm{m}$ as representative near-field, upper-near-field, and far-field operating points, whereas $100~\mathrm{m}$ is well
beyond the Rayleigh distance for any curvature and serves as a far-field baseline representative
of typical low-altitude links.

\begin{figure}[t!]
	\begin{centering}
		\includegraphics[width=0.48\textwidth]{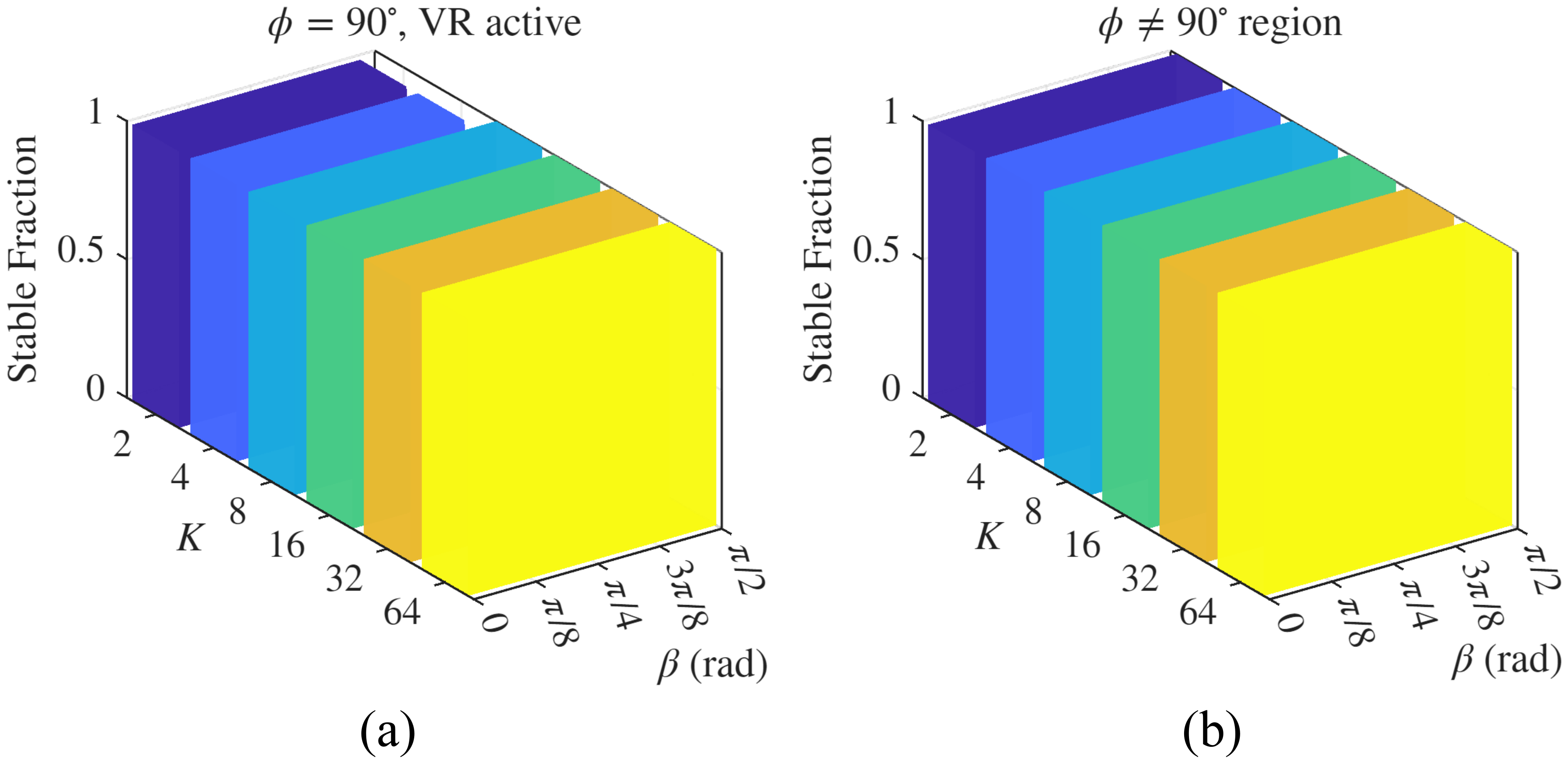}
        \captionsetup{font=footnotesize, labelsep=period}
		\caption{Local stable subarray fraction as a function of $\beta$ (rad) for 1D HoloCuRA with $N=128$, $L=0.32~\mathrm{m}$, and $r=2.0~\mathrm{m}$: (a) $\phi=90^\circ$ (VR active) and (b) the $\phi\neq 90^\circ$ region, both plotted versus the number of subarrays $K\in\{2,4,8,16,32,64\}$. Values closer to 1 indicate that a larger fraction of subarrays remains stable for the corresponding curvature/partition setting.}
\label{fig_localsnsula}
	\end{centering}
\end{figure}

\begin{figure}[t!]
	\begin{centering}
		\includegraphics[width=0.495\textwidth]{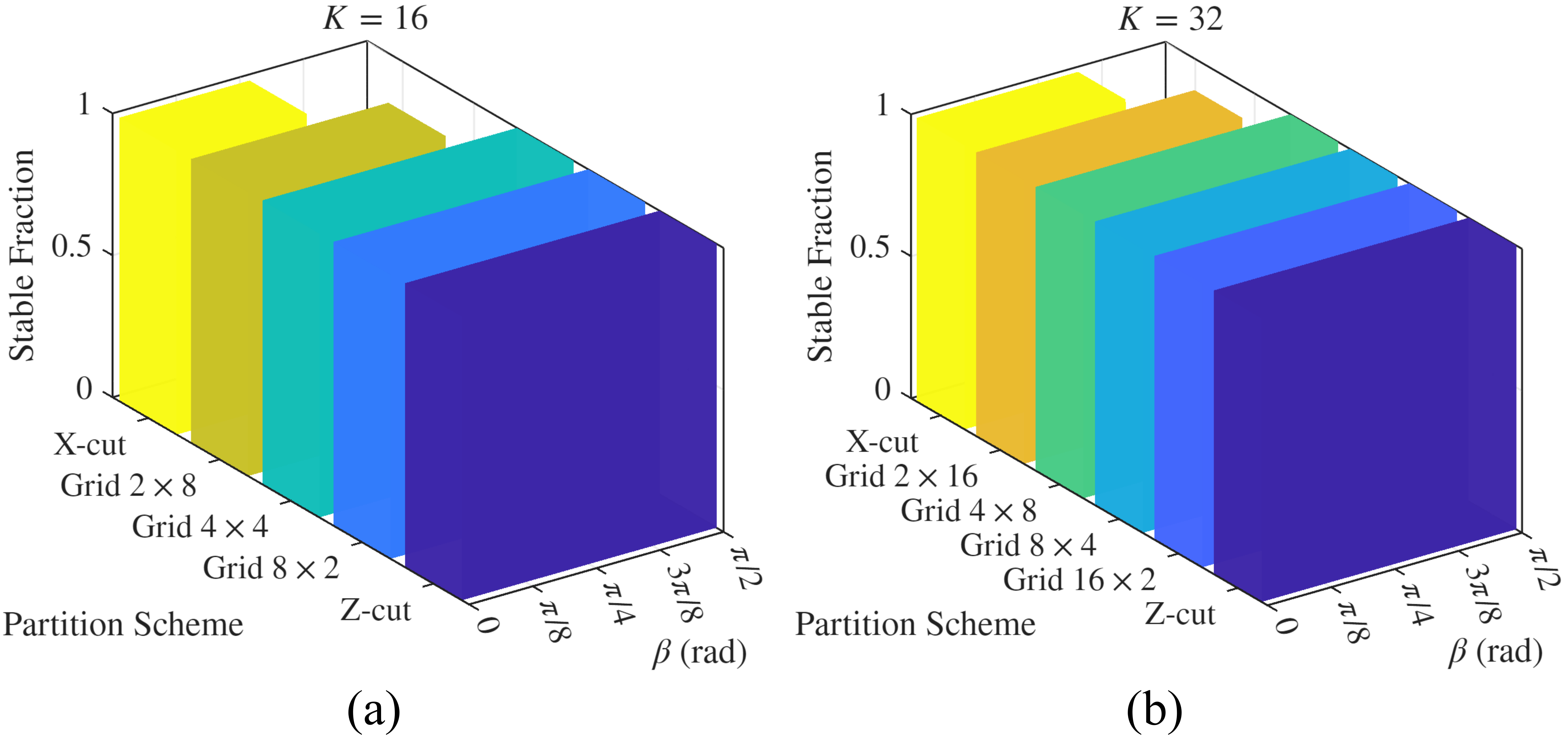}
        \captionsetup{font=footnotesize, labelsep=period}
		\caption{Local stable subarray fraction as a function of $\beta$ (rad) for 2D HoloCuRA local-SnS at $r=2.0~\mathrm{m}$ for (a) $K=16$ and (b) $K=32$, comparing different partition schemes (X-cut, Z-cut, and 2D grid partitions with the indicated grid sizes). Values closer to 1 indicate that a larger fraction of subarrays remains stable for the corresponding curvature/partition setting.}
\label{fig_localsnsura}
	\end{centering}
\end{figure}

\subsubsection{Admissible Partition Selection}

Fig.~\ref{fig_localsnsula} and Fig.~\ref{fig_localsnsura} show that, in 1D HoloCuRA, local SnS is dominated by VR-induced self-occlusion, whereas curvature alone causes only mild degradation away from the edge-on azimuth. In 2D HoloCuRA, increasing $K$ enlarges the admissible region because smaller subarrays span shorter VR gradients and weaker near-field power variation. The contrast between X-cut and Z-cut reveals that the dominant visibility variation is along the curved $z$-dimension. Here, X-cut partitions the aperture along $x$, yielding tiles that are narrow in $x$ but long in $z$, whereas Z-cut partitions along $z$, yielding tiles that are thin in $z$ but wide in $x$. Consequently, X-cut is more likely to mix rows with markedly different visibility states within the same subarray, producing larger intra-subarray power variation, whereas Z-cut better conforms to the underlying visibility structure and is therefore more locally stationary.

Guided by the above observations, we retain only those partition scales whose subarray footprint remains locally admissible under the LoS visibility variation, leading to $K_{\mathrm{ULA}}\in\{8,16\}$ for 1D HoloCuRA and $K_{\mathrm{URA}}\in\{16,32\}$ for 2D HoloCuRA. For 2D HoloCuRA, we further retain both Z-cut and representative 2D grids, because they probe two physically distinct ways in which curvature-induced visibility and propagation variation are distributed over the aperture: slicing along the dominant $z$-directed visibility gradient and compact two-dimensional tiling over the aperture surface. Concretely, for $K=16$ we compare Grid $16\times1$ with Grid $4\times4$, and for $K=32$ we compare Grid $32\times1$ with Grid $8\times4$. The resulting partitions contain roughly $512$--$1024$ elements per subarray, so that each subarray remains at a moderate and physically meaningful aperture scale rather than an arbitrarily fine discretization. 
Hence, subarray partitioning should not be interpreted as a free post-processing choice, nor does it create SnS; instead, it defines the aperture-domain observation scale at which the underlying geometry- and propagation-induced non-stationarity is represented.

\subsection{Array-Domain SnS under LoS}

After selecting admissible partitions, we evaluate full-aperture SnS by aggregating the pairwise PoVi-CMD defined in Section III. Local and array-domain SnS play different roles: the former determines whether a candidate partition remains locally admissible, whereas the latter quantifies the statistical disparity among subarrays over the full aperture. For each sampled direction $(\theta,\phi)$, let $d_{\mathrm{PoVi}}^{k\ell}$ denote the VR-aware pairwise discrepancy between subarrays $k$ and $\ell$, as defined in \eqref{eq:gamma}--\eqref{eq:dPoVi}. We then define the mean array-domain SnS level as
\begin{equation}
d_{\mathrm{PoVi}}^{\mathrm{mean}}(\theta,\phi)
=
\frac{2}{K(K-1)}
\sum_{k<\ell} d_{\mathrm{PoVi}}^{k\ell}(\theta,\phi),
\end{equation}
and map $d_{\mathrm{PoVi}}^{\mathrm{mean}}(\theta,\phi)$ over the sampled half-space. We emphasize that the retained $K$ values are not introduced to create SnS, but to probe the same geometry- and propagation-induced non-stationarity at different admissible observation scales.

\begin{figure}[t!]
	\centering
	\includegraphics[width=0.485\textwidth]{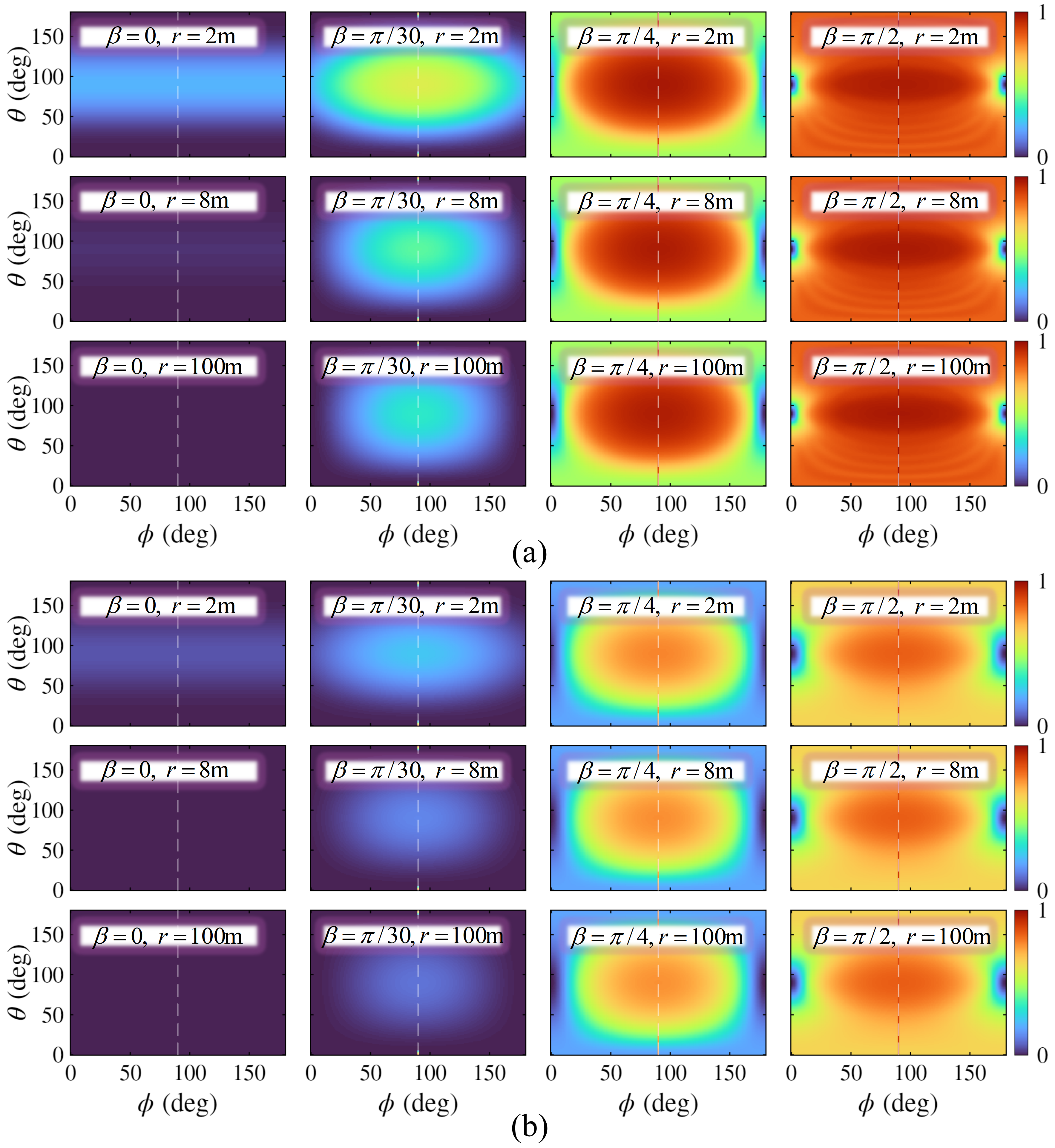}
    \captionsetup{font=footnotesize, labelsep=period}
    \caption{Heatmaps of the half-space-averaged array-domain metric $d_{\mathrm{PoVi}}$ (color scale) for 1D HoloCuRA. For each direction $(\theta,\phi)$, the shown value is $d_{\mathrm{PoVi}}$ averaged over all $K(K-1)/2$ subarray pairs using a Z-cut partition. Columns correspond to the array curvature $\beta\in\{0,\pi/30,\pi/4,\pi/2\}$, and rows correspond to the link distance $r\in\{2,8,100\}\,\mathrm{m}$. (a) $K=8$ and (b) $K=16$. The vertical dashed line marks $\phi=90^\circ$. Results are shown for $d_{\mathrm{el}}=\lambda/4$; the corresponding $d_{\mathrm{el}}=\lambda/2$ patterns are visually almost indistinguishable, and additional checks with $d_{\mathrm{el}}=\lambda/16$ lead to the same qualitative trends.}
\label{ULAheatmap1}
\end{figure}

\subsubsection{Near-to-Far-Field SnS of 1D HoloCuRA}

Fig.~\ref{ULAheatmap1} shows $d_{\mathrm{PoVi}}^{\mathrm{mean}}(\theta,\phi)$ for multiple ranges and curvatures. As $\beta$ increases, the high-SnS region broadens, while the dominant ridge remains centered near $\theta\approx 90^\circ$. For mild curvature, SnS decreases with $r$, which is consistent with the weakening of spherical-wave amplitude and phase gradients as the link moves toward the far field. For larger $\beta$, the dependence on $r$ becomes weaker, because the mismatch is then dominated by curvature-induced boresight divergence together with persistent VR-driven power imbalance.

The SnS heatmaps obtained with $d_{\mathrm{el}}=\lambda/2$ and $d_{\mathrm{el}}=\lambda/4$ are visually almost indistinguishable, so only the $d_{\mathrm{el}}=\lambda/4$ results are shown here for brevity. Additional checks with $d_{\mathrm{el}}=\lambda/16$ yield the same qualitative patterns, which indicates that the dominant SnS behavior is governed by the underlying aperture geometry and propagation rather than by the sampling grid itself. In this sense, dense holographic sampling does not create SnS, but suppresses discretization artifacts and makes its continuous-aperture origin more faithfully observable.

\begin{figure}[t!]
	\centering
	\includegraphics[width=0.48\textwidth]{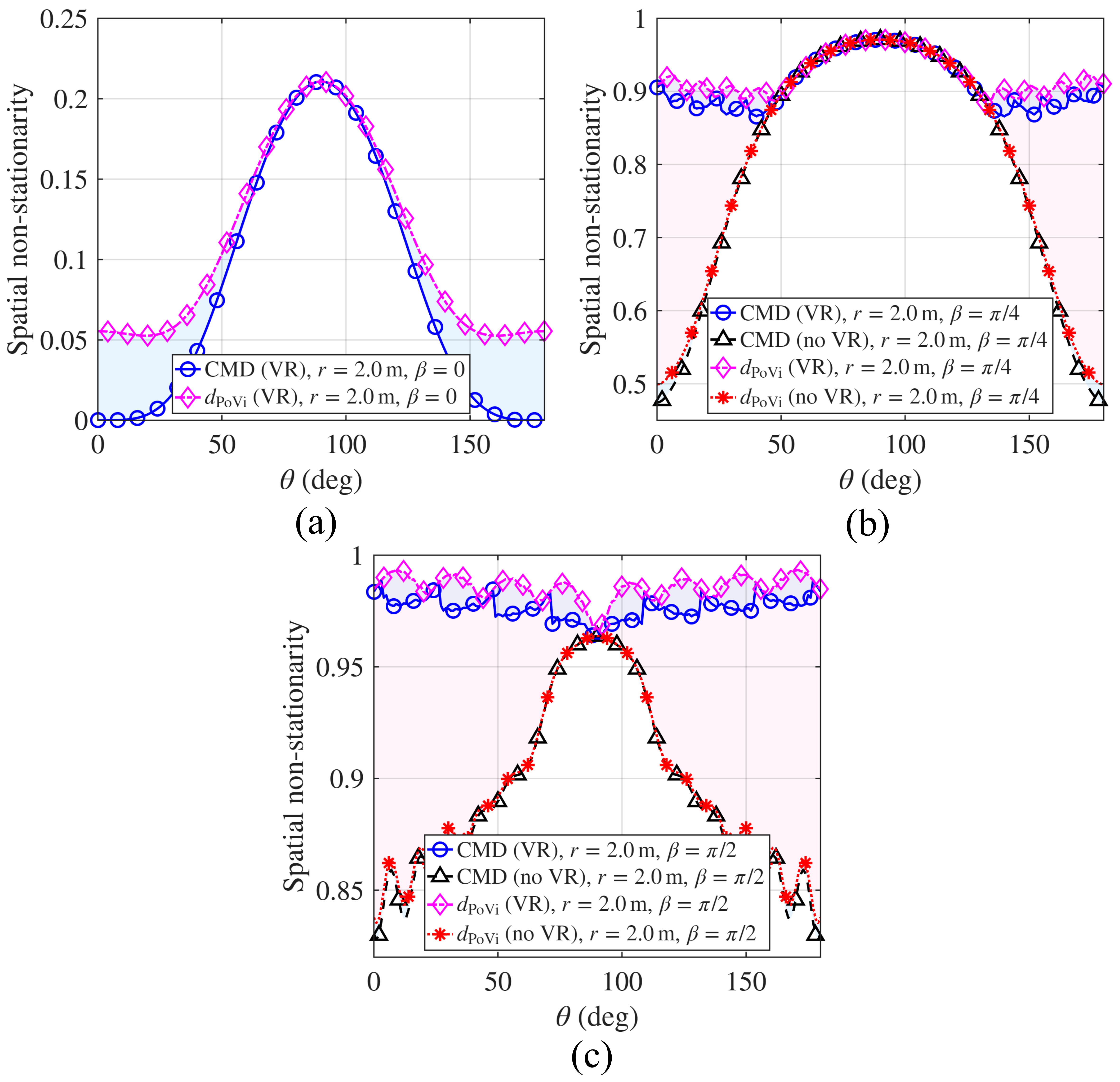}
    \captionsetup{font=footnotesize, labelsep=period}
    \caption{SnS versus elevation angle $\theta$ for 1D HoloCuRA at azimuth $\phi=\pi/2$ and range $r=2\,\mathrm{m}$. (a)--(c) correspond to array curvature $\beta=0$, $\beta=\pi/4$, and $\beta=\pi/2$, respectively. Two SnS measures are compared: the conventional CMD metric and the proposed $d_{\mathrm{PoVi}}$. For $\beta\in\{\pi/4,\pi/2\}$, results with and without VR modeling are both shown; for $\beta=0$, all elements are visible and only the VR-aware curves are plotted. The shown results use $d_{\mathrm{el}}=\lambda/4$; the corresponding $d_{\mathrm{el}}=\lambda/2$ and $d_{\mathrm{el}}=\lambda/16$ curves are visually almost indistinguishable.}
\label{fig:ula_heatmapcut}
\end{figure}

To sharpen the angular interpretation, Fig.~\ref{fig:ula_heatmapcut} plots $\theta$-cuts at $\phi=\pi/2$. As $\beta$ increases, the low-SnS trough contracts and the profile evolves into a high plateau with ripples, consistent with alternating inter-subarray alignment and misalignment compounded by VR on/off transitions. Unlike CMD, which is invariant to scalar power scaling, $d_{\mathrm{PoVi}}$ remains sensitive to VR-induced power imbalance through its power-balancing term and therefore gives a stronger and more range-robust indication of LoS SnS.

\subsubsection{Near-to-Far-Field SnS of 2D HoloCuRA}

\begin{figure*}[htbp]
\centerline{\includegraphics[width=0.99\linewidth]{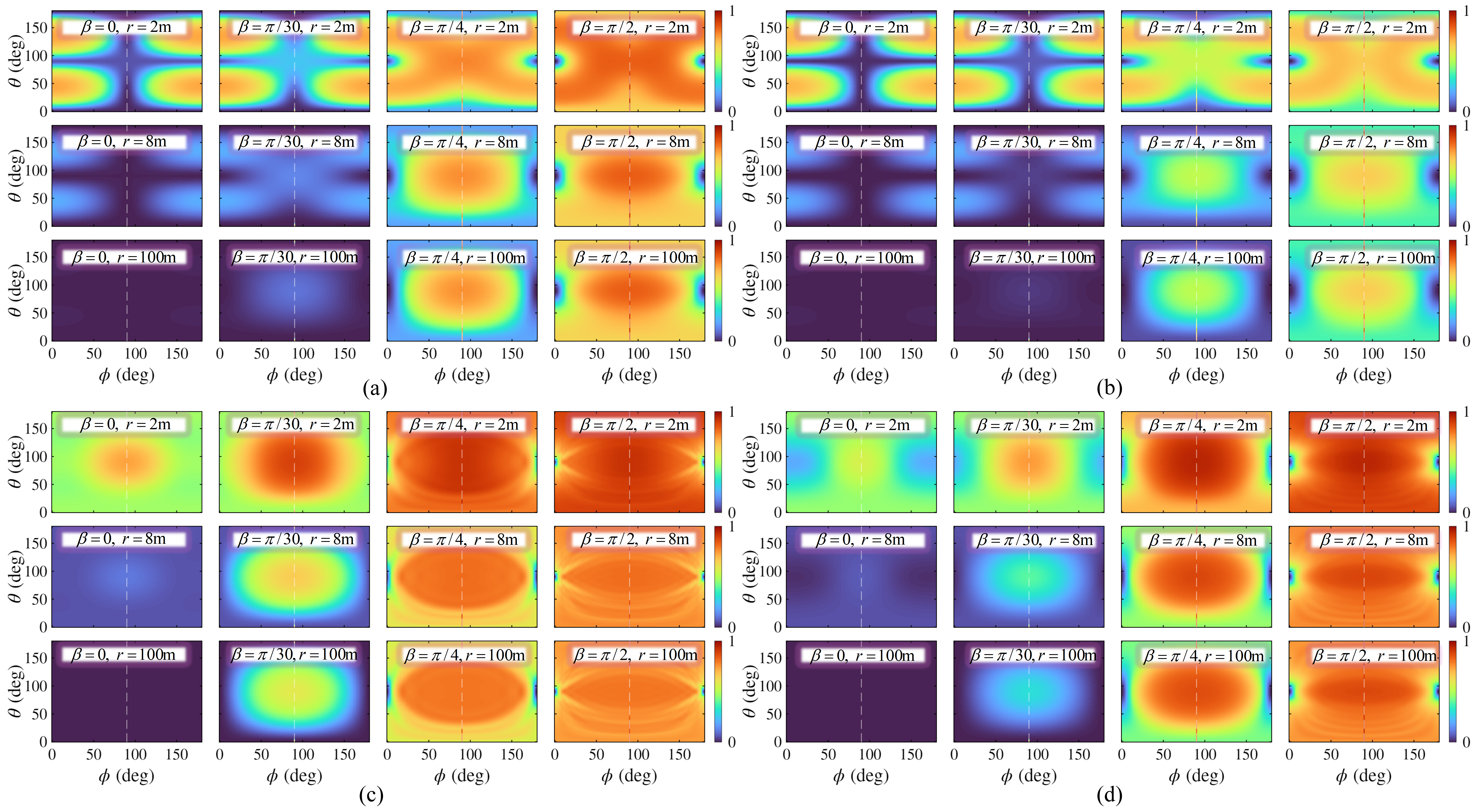}}
\captionsetup{font=footnotesize, labelsep=period}
\caption{Heatmaps of the half-space-averaged array-domain metric $d_{\mathrm{PoVi}}$ (color scale) for 2D HoloCuRA as a function of direction $(\theta,\phi)$ (in degrees). (a) and (c) show the case $K=16$, using a Z-cut partition (Grid $16\times1$, tile $8\times128$) and a 2D grid partition (Grid $4\times4$, tile $32\times32$), respectively. (b) and (d) show the case $K=32$, using a Z-cut partition (Grid $32\times1$, tile $4\times128$) and a 2D grid partition (Grid $8\times4$, tile $\approx16\times32$ or $32\times16$), respectively. The vertical dashed line marks $\phi=90^\circ$. Results are shown for $d_{\mathrm{el}}=\lambda/4$; the corresponding $d_{\mathrm{el}}=\lambda/2$ patterns are visually almost indistinguishable, and additional checks with $d_{\mathrm{el}}=\lambda/16$ lead to the same qualitative trends.}
\label{uraheatmap2}
\end{figure*}

Fig.~\ref{uraheatmap2} reports the array-domain SnS heatmaps of 2D HoloCuRA under the selected admissible partition schemes. For the Z-cut partition at short range ($r=2~\mathrm{m}$), a distinctive four-lobe pattern appears. This structure is caused by the spherical-wave phase variation along the $x$-dimension, which contains both a linear tilt term and a quadratic curvature term, and whose interplay produces symmetric correlation valleys across subarrays. As $r$ increases, the quadratic contribution decays on the order of $1/r$, so the lobes are gradually smoothed and the SnS contours become more circular.

For the representative 2D grids, the partition resolves both the $z$-directed visibility variation and the $x$-directed phase variation over the aperture. Under stronger curvature, the visibility boundary intersects more tiles, which creates partially visible subarrays and richer angular textures than in the Z-cut case. Across both 1D and 2D HoloCuRA, the array-domain SnS increases with curvature, whereas increasing $K$ mainly changes the observation scale: smaller subarrays are naturally more similar, but the same geometry- and propagation-induced trends remain.

As in the 1D case, the heatmaps obtained with $d_{\mathrm{el}}=\lambda/2$ and $d_{\mathrm{el}}=\lambda/4$ are visually almost indistinguishable, so only the $d_{\mathrm{el}}=\lambda/4$ results are shown here. Additional checks with $d_{\mathrm{el}}=\lambda/16$ again lead to the same qualitative patterns. This shows that, for a fixed physical aperture, reducing $d_{\mathrm{el}}$ mainly refines the sampling of the same continuous field rather than reshaping the underlying SnS pattern. By contrast, increasing $K$ changes the subarray footprint at which that field is observed. Hence, the weakening of the averaged SnS level at larger $K$ should be interpreted as a scale effect rather than as the disappearance of the underlying physical non-stationarity. This distinction is precisely where holographic sampling is useful: once the aperture is densely sampled, the observed SnS reflects the aperture-domain geometry and propagation more than the sampling grid itself.

Despite these common trends, 1D and 2D HoloCuRA remain fundamentally different. The 2D aperture preserves genuine azimuth--elevation discrimination and therefore supports a richer spatial organization of SnS than its 1D counterpart.

\subsection{Spatial DoF under LoS-Dominant Sparse Propagation}

\subsubsection{LoS Reference and CDL-D Baseline}

A pure single-ray LoS channel is rank-one and therefore serves as a geometric reference rather than a realistic modal baseline. To obtain a realistic, typical low-altitude, and reproducible LoS-type baseline beyond a single-ray
specular model, and to avoid the scenario dependence and limited transferability of
geometry-specific ray tracing, we adopt the 3GPP clustered delay line (CDL) model~\cite{3gpp_tr38901}. Here we use the standardized 3GPP CDL-D channel model, which includes a dominant specular component together with a limited number of clusters and finite angular spreads. This provides a sparse-support reference that remains consistent with low-altitude LoS-dominant propagation.

\subsubsection{Eigen-Spectrum and DoF Comparison}

\begin{figure}[t!]
 \begin{centering}
  \includegraphics[width=0.485\textwidth]{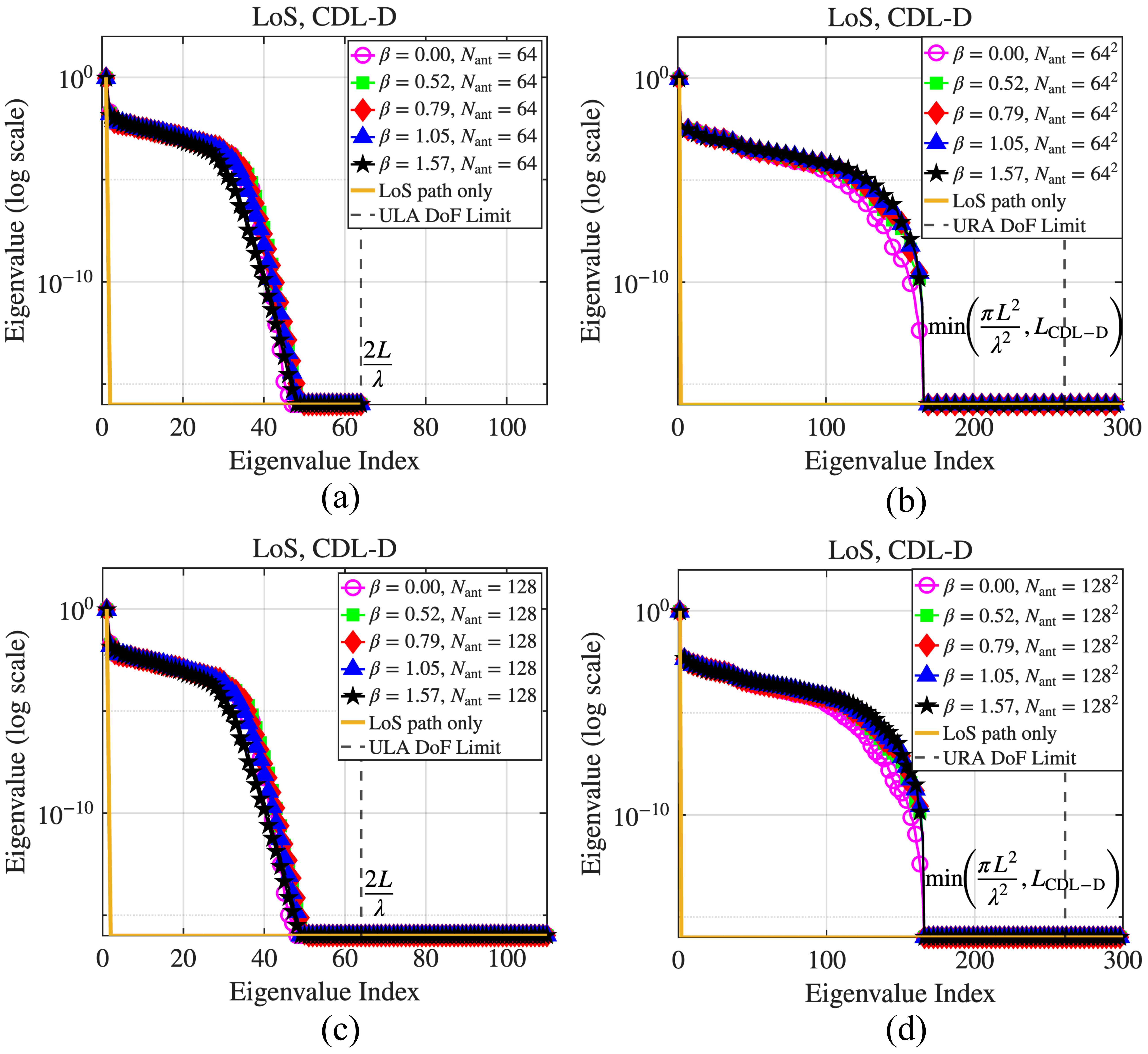}
        \captionsetup{font=footnotesize, labelsep=period}
  \caption{Sorted eigenvalue spectra of the HMIMO spatial correlation matrix under a LoS link and the 3GPP CDL-D channel model for HoloCuRA. (a) 1D HoloCuRA with half-wavelength inter-element spacing $d=\lambda/2$ ($N_{\mathrm{ant}}=64$). (b) 2D HoloCuRA with $d=\lambda/2$ ($N_{\mathrm{ant}}=64^2$). (c) 1D HoloCuRA with quarter-wavelength spacing $d=\lambda/4$ ($N_{\mathrm{ant}}=128$). (d) 2D HoloCuRA with $d=\lambda/4$ ($N_{\mathrm{ant}}=128^2$). Eigenvalues are ordered in decreasing magnitude. Vertical dashed lines indicate the HMIMO spatial DoF limits: $2L/\lambda$ for the 1D array and $\min(\pi L^2/\lambda^2,\, L_{\mathrm{CDL-D}})$ for the 2D array. Here $L_{\mathrm{CDL-D}}$ denotes the effective DoF upper bound imposed by the 3GPP CDL-D multipath channel model.}
\label{fig_losdof}
 \end{centering}
\end{figure}

As shown in Fig.~\ref{fig_losdof}, under the CDL-D model both 1D and 2D HoloCuRA exhibit eigen-spectra that are strongly concentrated at low orders and decay rapidly thereafter, which indicates that the effective DoF is activated only over a limited set of dominant spatial modes under LoS-dominant sparse propagation. Relative to 1D HoloCuRA, the 2D aperture sustains a longer tail and a later spectral knee, because the additional aperture dimension preserves extra spatial discrimination under the same sparse angular support. By contrast, the single-ray LoS reference remains essentially rank-one and therefore serves mainly as a geometric lower-complexity baseline.

The impact of curvature $\beta$ is secondary compared with the overall low-rank structure, but it is not negligible. Its effect appears primarily in the high-order tail, where curvature perturbs the array manifold through changes in local aperture orientation and projected propagation support. This perturbation is more visible for 2D HoloCuRA, since the extra aperture dimension allows curvature-induced changes to be distributed over a richer set of weak modes. In 1D HoloCuRA, however, the array remains intrinsically one-dimensional, so under CDL-D's LoS-dominant excitation the higher-order tail need not evolve monotonically with $\beta$; instead, curvature mainly reshapes a small number of weakly excited modes without altering the overall low-order concentration.

Comparing the $d_{\mathrm{el}}=\lambda/2$ cases in Fig.~\ref{fig_losdof}(a)--(b) with the $d_{\mathrm{el}}=\lambda/4$ cases in Fig.~\ref{fig_losdof}(c)--(d) further shows that denser holographic sampling refines the spectral representation but does not materially shift the dominant spectral knee or the effective DoF. For 1D HoloCuRA, the $\lambda/2$ sampling exhibits an almost hard truncation after the knee, whereas $\lambda/4$ mainly reveals additional near-zero modes beyond the effective cutoff without creating new significant ones. For 2D HoloCuRA, the main difference is again spectral resolution: the denser sampling provides more eigenvalue samples over the same DoF region and yields a clearer depiction of the roll-off near the cutoff, while the cutoff location itself is essentially preserved.

Taken together, these results show that in LoS-dominant sparse propagation the effective DoF is governed primarily by aperture dimensionality and propagation support, whereas curvature mainly acts as a secondary perturbation on the eigenvalue tail. In this sense, holographic sampling does not increase the physical DoF of a fixed aperture, but improves the fidelity with which the underlying eigen-spectrum is resolved and interpreted.

\subsection{Joint LoS Insights on SnS and DoF}

Under LoS, the half-space heatmaps of $d_{\mathrm{PoVi}}$ show where different subarrays still exhibit similar statistics and where they begin to differ over $(\theta,\phi)$, $r$, and $\beta$. In this sense, the proposed SnS framework does more than assign a scalar non-stationarity level: by combining VR-aware subarray partitioning with a local-to-global screening procedure, it turns curvature-induced spatial heterogeneity into an explicit aperture-domain characterization that can be resolved over direction and range.

The LoS results further show that array-domain SnS is not a single-mechanism effect. Rather, it is jointly produced by spherical-wave propagation, curvature-dependent, and VR-induced visibility interruption. These mechanisms act on different but coupled scales: spherical-wave gradients dominate in the radiating near field, curvature reshapes the aperture response through local orientation changes, and VR introduces abrupt power imbalance through self-occlusion. Their combined action makes the per-element and per-subarray responses strongly location dependent, especially at short range.

Viewed together with the DoF results, this leads to a clear design implication for HoloCuRA. 
One is that local SnS screening is indispensable for selecting admissible subarray sizes and orientations under self-occlusion and near-field effects, thereby enabling stable subarray level processing and calibration. 
The other is that the pronounced eigenvalue concentration under CDL-D shows diminishing DoF returns in directionally sparse environments. 
Therefore, under low-altitude LoS conditions, array-domain SnS directly indicates whether curvature- and VR-induced aperture variation is strong enough to support subarray-wise operation: large SnS favors partitioned use of HoloCuRA, whereas small SnS suggests that the aperture should be used more effectively as a single coherent surface.

\section{NLoS Spatial Characterization Under Non-Isotropic Scattering}
\label{sec:nlos}

To extend the LoS geometry-dominated analysis to a statistically grounded NLoS setting, we retain the $d_{\mathrm{PoVi}}$ framework so that SnS is quantified on a common scale across propagation environments. In NLoS, however, non-stationarity is no longer governed by explicit self-occlusion alone, but by non-isotropic angular support and curvature-dependent spatial correlation. This section therefore examines how curvature reshapes the angular support, how array domain SnS differs from spatial WSS, and how these effects relate to the effective DoF under CDL-A.

\subsection{Non-Isotropic Scattering Model and SnS Indicators}

\subsubsection{Curvature-Dependent Angular Support}

Under NLoS non-isotropic scattering, the spatial correlation is computed from the discrete three-dimensional angular power spectrum specified by the CDL model. Unlike the LoS case, where backside scattering is neglected and the effective support reduces to a $\beta$-independent front half-space, NLoS propagation admits non-negligible backside contributions, which produce a $\beta$-dependent coverage extension. Specifically, bending rotates the local surface normal of each element over $[-\beta,\beta]$, enlarging the union of forward-visible directions. The additional backside support is thus confined to the two elevation caps
\begin{equation}
\theta\in[0,\beta]\cup[\pi-\beta,\pi],
\end{equation}
while spanning the full azimuth range $\phi\in[0,2\pi]$.

\subsubsection{SnS Mean and Variance Versus Subarray Separation}

A second distinction from LoS is that, under NLoS, the channel is characterized by multi-cluster angular support with distinct AoA/AoD statistics under the standard far-field assumption~\cite{3gpp_tr38901}. In this setting, pointwise directional comparison is less informative than a separation-based array-domain summary. We therefore define two SnS indicators over the subarray separation $s=|k-\ell|$: the mean discrepancy
\begin{equation}
\bar d(s;\beta)=\frac{1}{K-s}\sum_{k=1}^{K-s} d_{\mathrm{PoVi}}(k,k+s;\beta),
\end{equation}
and the corresponding same-separation variance
\begin{equation}
\mathrm{Var}_{s}(\beta)=\mathrm{Var}_{k=1,\dots,K-s}\!\left[d_{\mathrm{PoVi}}(k,k+s;\beta)\right].
\end{equation}

If the array-domain statistics were spatially WSS along the aperture, they would be invariant to index translation, which implies $\mathrm{Var}_{s}(\beta)=0$ for every $s$. Hence, $\bar d(s;\beta)$ measures the overall SnS strength, whereas $\mathrm{Var}_{s}(\beta)$ quantifies the departure from spatial WSS.

\subsection{CDL-A Baseline and Evaluation Setup}

To obtain a standardized and reproducible NLoS baseline, we adopt CDL-A from 3GPP TR~38.901~\cite{3gpp_tr38901}. CDL-A features relatively narrow three-dimensional angular spreads, e.g., $c_{\mathrm{ASA}}=11^\circ$ and $c_{\mathrm{ZSA}}=3^\circ$, which is consistent with directionally sparse low-altitude NLoS links dominated by a limited number of strong clusters. The same admissible partition scales selected in the LoS section are retained here so that the comparison across propagation environments is performed under a common subarray architecture.

\begin{figure}[t!]
 \begin{centering}
  \includegraphics[width=0.50\textwidth]{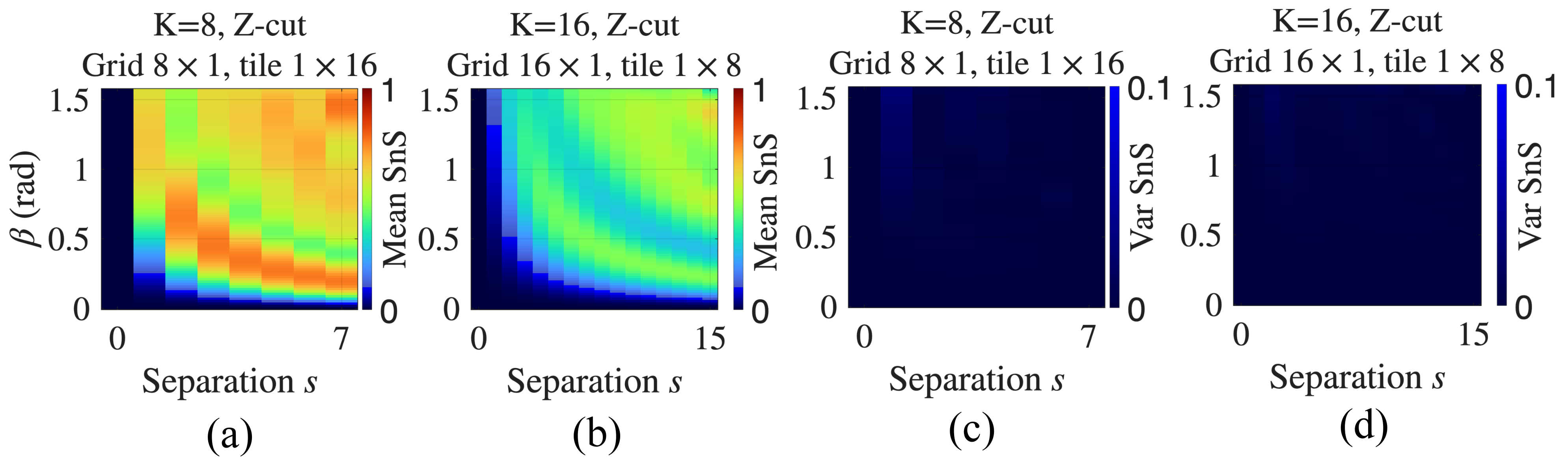}
        \captionsetup{font=footnotesize, labelsep=period}
  \caption{Z-cut maps of SnS for 1D HoloCuRA in the 3GPP CDL-A (NLoS) channel. The mean SnS is shown in (a)--(b) and the SnS variance in (c)--(d), as functions of the subarray separation $s$ and curvature $\beta$ (rad). Panels (a),(c) correspond to $K=8$, whereas (b),(d) correspond to $K=16$ (the grid/tile settings used for each case are indicated above each subplot).}
\label{fig_snsnlosula}
 \end{centering}
\end{figure}

\begin{figure}[t!]
 \begin{centering}
  \includegraphics[width=0.50\textwidth]{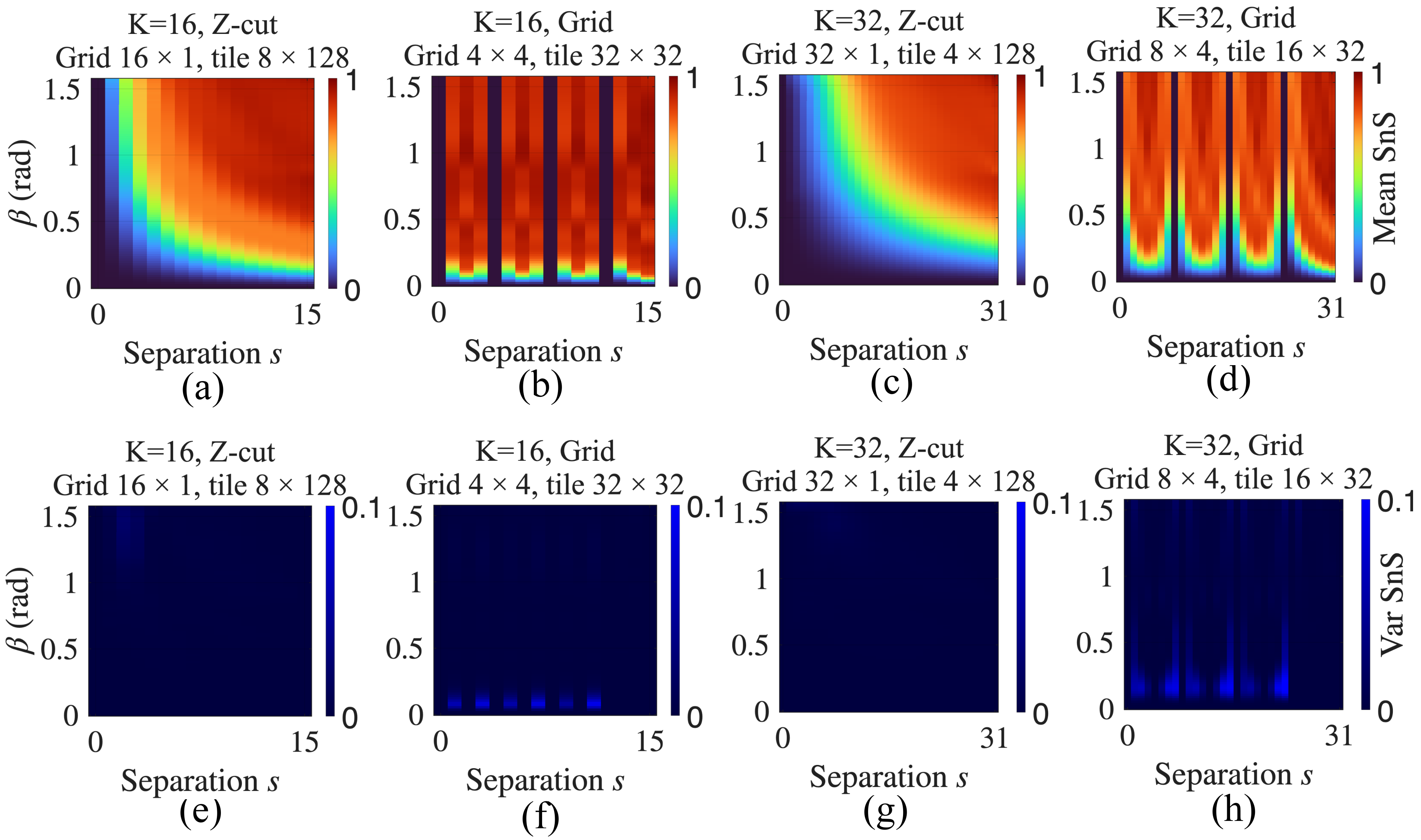}
\captionsetup{font=footnotesize, labelsep=period}
  \caption{Comparison of Z-cut and grid-based evaluations of SnS for HoloCuRA under the 3GPP CDL-A channel. Heatmaps of the \emph{mean} SnS are shown in (a)--(d), and those of the SnS \emph{variance} in (e)--(h), versus the subarray separation $s$ and curvature $\beta$ (rad). Results for $K=16$ are given in (a),(b),(e),(f), and results for $K=32$ are given in (c),(d),(g),(h). For each $K$, the Z-cut maps are shown in (a),(c),(e),(g), while the corresponding grid-based maps are shown in (b),(d),(f),(h). Grid/tile configurations are indicated above each subplot; colorbars report the mean SnS (top row) and variance (bottom row).}
\label{fig_snsnlosura}
 \end{centering}
\end{figure}

\subsection{SnS Behavior Under CDL-A}

\subsubsection{1D HoloCuRA}

For 1D HoloCuRA, Fig.~\ref{fig_snsnlosula} shows that $\bar d(s;\beta)$ exhibits visible ripples over $(s,\beta)$, reflecting the discrete angular support of CDL-A: with only a finite number of dominant clusters and rays, small geometry-induced phase reweightings can produce constructive and destuctive changes in the averaged SnS. Increasing $K$ reduces the overall SnS level because smaller subarrays are naturally more similar. Meanwhile, $\mathrm{Var}_{s}(\beta)$ remains nearly negligible, indicating that in this 1D setting the discrepancy is governed primarily by the subarray separation $s$ rather than by the absolute subarray location.

\subsubsection{2D HoloCuRA}

For 2D HoloCuRA, Fig.~\ref{fig_snsnlosura} shows that $\bar d(s;\beta)$ generally increases with both curvature $\beta$ and separation $s$. Under Z-cut, the mean-SnS surface is mildly wavy yet remains approximately monotone, consistent with a dominant variation direction along the aperture. Under grid partitioning, the same $s$ can mix subarray pairs from different physical directions. Because curvature influences different directions unequally, the resulting heatmap exhibits regular stripe-like patterns. Since curvature does not affect all directions equally, pairs aligned with the more sensitive direction, where element orientation and cluster visibility vary more strongly, exhibit larger SnS, whereas pairs aligned mainly with less sensitive directions exhibit smaller SnS. Such stripes are not artifacts, but a physically meaningful reflection of the directional nature of the 2D curved aperture.

Despite these texture differences, the main trends are consistent across CDL-A and across both 1D and 2D HoloCuRA: (i) SnS remains weak as $\beta\rightarrow 0$, corresponding to nearly linear or planar apertures; (ii) SnS decreases as $K$ increases, because the same non-stationarity is observed at a finer subarray scale; and (iii) SnS increases with both curvature and subarray separation. Moreover, the low level of $\mathrm{Var}_{s}(\beta)$ indicates that CDL-A, although non-isotropic, still concentrates most of its variation into a limited number of dominant directions rather than producing a strongly position-dependent random result over the aperture.

\begin{figure}[t!]
 \begin{centering}
  \includegraphics[width=0.485\textwidth]{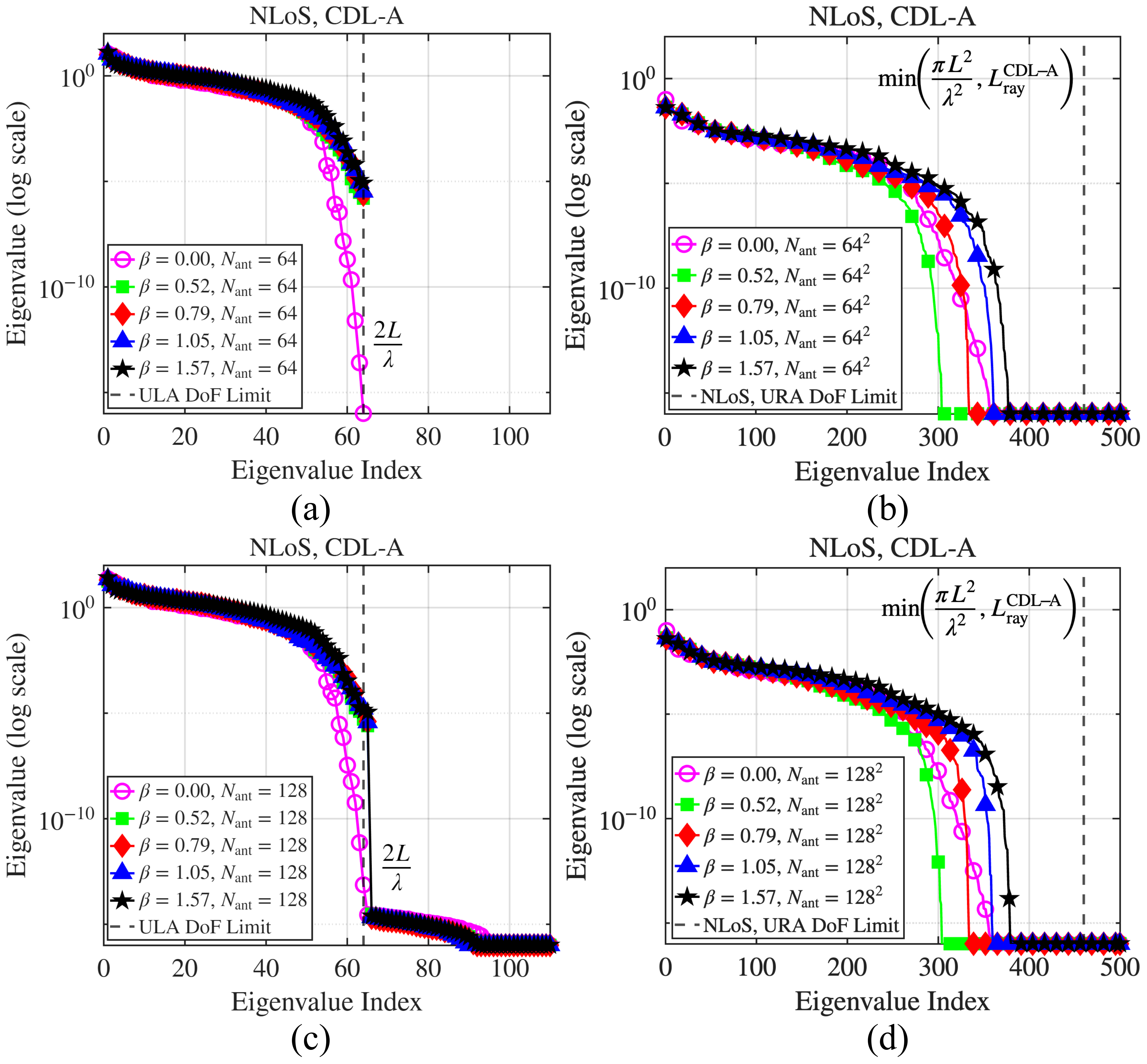}
        \captionsetup{font=footnotesize, labelsep=period}
        \caption{Sorted eigenvalue spectra of the spatial correlation matrix for HoloCuRA under the 3GPP CDL-A NLoS channel. (a),(c) 1D HoloCuRA with $N_{\mathrm{ant}}=64$ and $N_{\mathrm{ant}}=128$, respectively; (b),(d) 2D HoloCuRA with $N_{\mathrm{ant}}=64^2$ and $N_{\mathrm{ant}}=128^2$, respectively. Eigenvalues are ordered in decreasing magnitude and plotted versus the eigenvalue index on a logarithmic scale. The vertical dashed lines indicate the spatial DoF limits: $2L/\lambda$ for the ULA and $\min(\pi L^2/\lambda^2,\, L_{\mathrm{ray}}^{\mathrm{CDL-A}})$ for the URA, where $L_{\mathrm{ray}}^{\mathrm{CDL-A}}$ denotes the DoF ceiling imposed by the finite number of paths in the CDL-A model.}
  \label{fig_dofnloscdla}
 \end{centering}
\end{figure}

\subsection{Spatial DoF Under CDL-A}

Fig.~\ref{fig_dofnloscdla} shows the eigenvalue spectra of the spatial correlation matrices for 1D and 2D HoloCuRA under CDL-A. In all cases, the spectra roll off well before the physical-aperture DoF bound, which indicates that the effective DoF is primarily scattering-limited in this NLoS scenario. As $\beta$ increases, the truncation point tends to move toward higher orders, although not strictly monotonically, reflecting the interaction between curvature-induced manifold perturbation and sparse angular support.

The comparison between $d_{\mathrm{el}}=\lambda/2$ and $d_{\mathrm{el}}=\lambda/4$ follows the same pattern as in the LoS case: denser holographic sampling refines the spectral representation but does not materially shift the dominant knee or the effective DoF. For 1D HoloCuRA, the finer sampling mainly reveals additional weak modes beyond the cutoff; for 2D HoloCuRA, it provides a denser sampling of the same roll-off region. Thus, under CDL-A the effective DoF remains controlled primarily by the available scattering support and aperture dimensionality, while curvature acts mainly as a secondary perturbation of the eigenvalue tail.

\subsection{Joint NLoS Insights on SnS and DoF}

Taken together, the NLoS results show that curvature affects HoloCuRA through two coupled mechanisms. First, by reshaping the angular support and the corresponding spatial correlation, it modifies the array-domain SnS and can make a global WSS approximation even under a standardized CDL baseline. Second, through the same support reshaping, it perturbs the eigen-spectrum and slightly redistributes the usable spatial modes, although the dominant DoF bottleneck remains the limited number of excited scattering directions.

Therefore, under NLoS non-isotropic scattering, the spatial behavior of HoloCuRA cannot be inferred from aperture size or sampling density alone. SnS still depends on how curvature interacts with the dominant scattering directions and how the aperture is partitioned, whereas DoF remains bounded primarily by the available propagation support. This confirms
that curvature, subarray geometry, and scattering structure
must be considered jointly when characterizing HoloCuRA
under realistic NLoS conditions.

\section{Isotropic Spatial Characterization: Closed-Form Correlation, SnS, and DoF}

Consistent with the holographic viewpoint of this paper, we model HoloCuRA as a quasi-continuous electromagnetic aperture and derive its spatial correlation under half-space isotropic scattering. Unlike the LoS and NLoS cases, where spatial heterogeneity is shaped by visibility interruption or non-isotropic angular support, isotropic scattering provides a reference in which the role of curvature becomes analytically transparent. This section therefore serves as the continuous-aperture baseline of the paper: it yields closed-form correlation expressions for 1D and 2D HoloCuRA, clarifies the effect of curvature-induced coverage extension, and provides a weak-SnS reference against which the stronger non-stationarity observed under LoS and NLoS can be interpreted.

\subsection{Continuous-Aperture Isotropic Correlation Model}

We begin from the spatial correlation matrix
\begin{equation}
\mathbf{R}=\mathbb{E}\!\left\{\mathbf{a}(\theta,\varphi)\mathbf{a}^{H}(\theta,\varphi)\right\},
\end{equation}
whose entries are angular integrals weighted by a scattering function $S(\theta,\varphi)$. If
$S(\theta,\varphi)$ is normalized over an integration domain $\mathcal{D}$, i.e.,
\begin{equation}
\iint_{\mathcal{D}}S(\theta,\varphi)\,d\Omega=1,
\end{equation}
then the diagonal entries satisfy $[R]_{m,m}=1$, and $\mathbf{R}$ is a correlation-coefficient matrix. Throughout this section we adopt half-space isotropic scattering over
$\mathcal{D}_{0}=[0,\pi]\times[0,\pi]$~\cite{Ref_SunTao2022},
\begin{equation}
S(\theta,\varphi)=\frac{\sin\theta}{2\pi},
\qquad
(\theta,\varphi)\in\mathcal{D}_{0},
\end{equation}
and use the normalized sinc function $\sinc(x)=\sin(\pi x)/(\pi x)$.

\subsubsection{1D HoloCuRA: Baseline Correlation and Coverage Extension}
\label{sec:Flexible_ula_corr}

Under $\mathcal{D}_0$ and the dominant zeroth-order approximation, the baseline correlation of 1D HoloCuRA depends only on the central-angle separation $\alpha_n-\alpha_m$ and admits the closed form~\cite{VTC2025}
\begin{equation}
\label{eq:ula_baseline_sinc_main}
[R_{0}]_{m,n}
\approx
\sinc\!\left(
\frac{4R}{\lambda}\,
\sin\!\left(\frac{\alpha_{n}-\alpha_{m}}{2}\right)
\right).
\end{equation}

Bending may additionally expose part of the backside azimuth support at extreme elevations. We model this coverage extension through
\begin{multline}
\label{eq:Omega_ext_ula_main}
\Omega_{\mathrm{ext}}(\beta)\triangleq
\big\{(\theta,\varphi):\ \theta\in[0,\beta]\cup[\pi-\beta,\pi],\\
\varphi\in[\pi,2\pi]\big\},
\end{multline}
while keeping $S(\theta,\varphi)$ unchanged. The total correlation can then be decomposed as
\begin{equation}
\label{eq:ula_total_decomp_main}
R_{m,n}(\beta)=R_{0,m,n}+R_{\mathrm{extra},m,n}(\beta),
\end{equation}
where the extension term reduces to the bounded one-dimensional correction
\begin{multline}
\label{eq:ula_extra_1d_main}
R_{\mathrm{extra},m,n}(\beta)
\approx
\int_{\cos\beta}^{1}
\cos\!\big(bu\cos c\big)\\
\times
J_{0}\!\Big(b\sqrt{1-u^{2}}\sin c\Big)\,du,
\end{multline}
with $u=\cos\theta$ and
\begin{IEEEeqnarray}{rCl}
\label{b,c}
b &=&
\frac{4\pi R}{\lambda}\,
\sin\!\left(\frac{\alpha_{n}-\alpha_{m}}{2}\right),\\
c &=&
\beta-\frac{\alpha_{m}+\alpha_{n}}{2}.
\end{IEEEeqnarray}

Because the angular domain is enlarged without re-normalizing $S(\theta,\varphi)$, the diagonal power increases to
\begin{equation}
\label{eq:ula_diag_main}
R_{m,m}(\beta)=
\iint_{\mathcal{D}_{0}\cup\Omega_{\mathrm{ext}}(\beta)}
\frac{\sin\theta}{2\pi}\,d\Omega
=2-\cos\beta.
\end{equation}

Accordingly, the normalized correlation-coefficient matrix is
\begin{equation}
\label{eq:ula_norm_main}
\tilde{R}_{m,n}(\beta)=\frac{R_{m,n}(\beta)}{2-\cos\beta}.
\end{equation}

\subsubsection{2D HoloCuRA: Baseline Correlation and Coverage Extension}
\label{sec:Flexible_ura_corr}

For 2D HoloCuRA, the same half-space isotropic model yields the baseline correlation between elements $(m,n)$ and $(m',n')$ as~\cite{VTC2025}
\begin{equation}
\label{eq:ura_baseline_main}
R^{(0)}_{(m,n),(m',n')}
\approx
\sinc\!\left(
\frac{2}{\lambda}\sqrt{D^{2}+E^{2}}
\right),
\end{equation}
where the geometric quantities $D$ and $E$ are
\begin{equation}
\label{eq:DE_main}
\begin{cases}
\scalebox{0.86}{$
\begin{aligned}
D \;&=\; \sqrt{\,(m - m')^2\,d^2 \;+\; 4\,R^2\,\sin^2\Bigl(\tfrac{\psi_n - \psi_{n'}}{2}\Bigr)\,\sin^2\Bigl(\beta - \tfrac{\psi_n + \psi_{n'}}{2}\Bigr)},\\[6pt]
E \;&=\; 2\,R \,\cos\!\Bigl(\beta - \tfrac{\psi_n + \psi_{n'}}{2}\Bigr)\,\sin\!\Bigl(\tfrac{\psi_n - \psi_{n'}}{2}\Bigr).
\end{aligned}
$}
\end{cases}
\end{equation}

Using the same coverage-extension region \eqref{eq:Omega_ext_ula_main}, the total correlation becomes
\begin{equation}
\label{eq:ura_total_main}
R_{(m,n),(m',n')}(\beta)
\approx
R^{(0)}_{(m,n),(m',n')}
+
R_{\mathrm{extra},(m,n),(m',n')}(\beta),
\end{equation}
where the extension term again reduces to a bounded one-dimensional integral,
\begin{equation}
\label{eq:ura_extra_main}
\begin{aligned}
R_{\mathrm{extra},(m,n),(m',n')}(\beta)
&\approx \int_{\cos\beta}^{1}
\cos\!\left(k_0Eu\right) \\
&\quad \cdot
J_{0}\!\Bigl(k_0D\sqrt{1-u^{2}}\Bigr)\,du .
\end{aligned}
\end{equation}
with $k_0=2\pi/\lambda$ and $u=\cos\theta$. As in the 1D case, the diagonal equals the total collected average power over $\mathcal{D}_0\cup\Omega_{\mathrm{ext}}(\beta)$ and is therefore given by $2-\cos\beta$. The normalized correlation-coefficient matrix is
\begin{equation}
\label{eq:ura_norm_main}
\tilde{R}_{(m,n),(m',n')}(\beta)
=
\frac{R_{(m,n),(m',n')}(\beta)}{2-\cos\beta}.
\end{equation}

Taken together, the 1D and 2D results show that under isotropic scattering the HoloCuRA correlation reduces to a sinc-type baseline plus a bounded one-dimensional correction. This is precisely where the holographic viewpoint is most useful: once the aperture is treated as densely sampled, the underlying continuous-aperture correlation kernel becomes both analytically tractable and physically interpretable.

\subsection{Correlation-Coefficient Matrices for 1D and 2D HoloCuRA}

The normalized matrices $\tilde{\mathbf R}^{\mathrm{iso}}(\beta)$ in \eqref{eq:ula_norm_main} and \eqref{eq:ura_norm_main} provide a unified second-order description of isotropic HoloCuRA. For any prescribed subarray partition with index sets $\{\mathcal I_k\}_{k=1}^{K}$, the corresponding subarray correlation blocks are obtained directly as
\[
\tilde{\mathbf R}^{\mathrm{iso}}(\beta)[\mathcal I_k,\mathcal I_\ell],
\]
so that the SnS metrics based on $d_{\mathrm{PoVi}}$ can be evaluated deterministically, without additional angular sampling or Monte-Carlo averaging. This direct block extraction is important because it connects the continuous-aperture correlation model to the same subarray-level framework used in the LoS and NLoS sections.

\subsection{SnS Under Isotropic Scattering}

Under isotropic scattering, angular averaging smooths out direction-specific irregularities and yields a near-stationary second-order reference. In the planar limit ($\beta=0$), the resulting matrix reduces to a Toeplitz structure for 1D HoloCuRA and to a block-Toeplitz-with-Toeplitz-blocks structure for 2D HoloCuRA. Consequently, for any equal-size contiguous or grid partition, equal-separation subarray pairs share the same or nearly the same second-order statistics, which implies a near-zero same-separation SnS variance and a mean SnS that depends primarily on the subarray displacement rather than on the absolute aperture location.

For $\beta>0$, bending introduces only a bounded coverage-extension correction and a smooth diagonal scaling by $2-\cos\beta$. Hence, curvature perturbs the isotropic correlation kernel only mildly and preserves near translation invariance over the aperture. In this sense, isotropic scattering provides a weak-SnS baseline: the observed non-stationarity remains small, and stronger SnS in the LoS and NLoS sections should therefore be interpreted as arising from visibility interruption or non-isotropic angular support rather than from the HoloCuRA alone.

\begin{figure}[t!]
	\begin{centering}
		\includegraphics[width=0.485\textwidth]{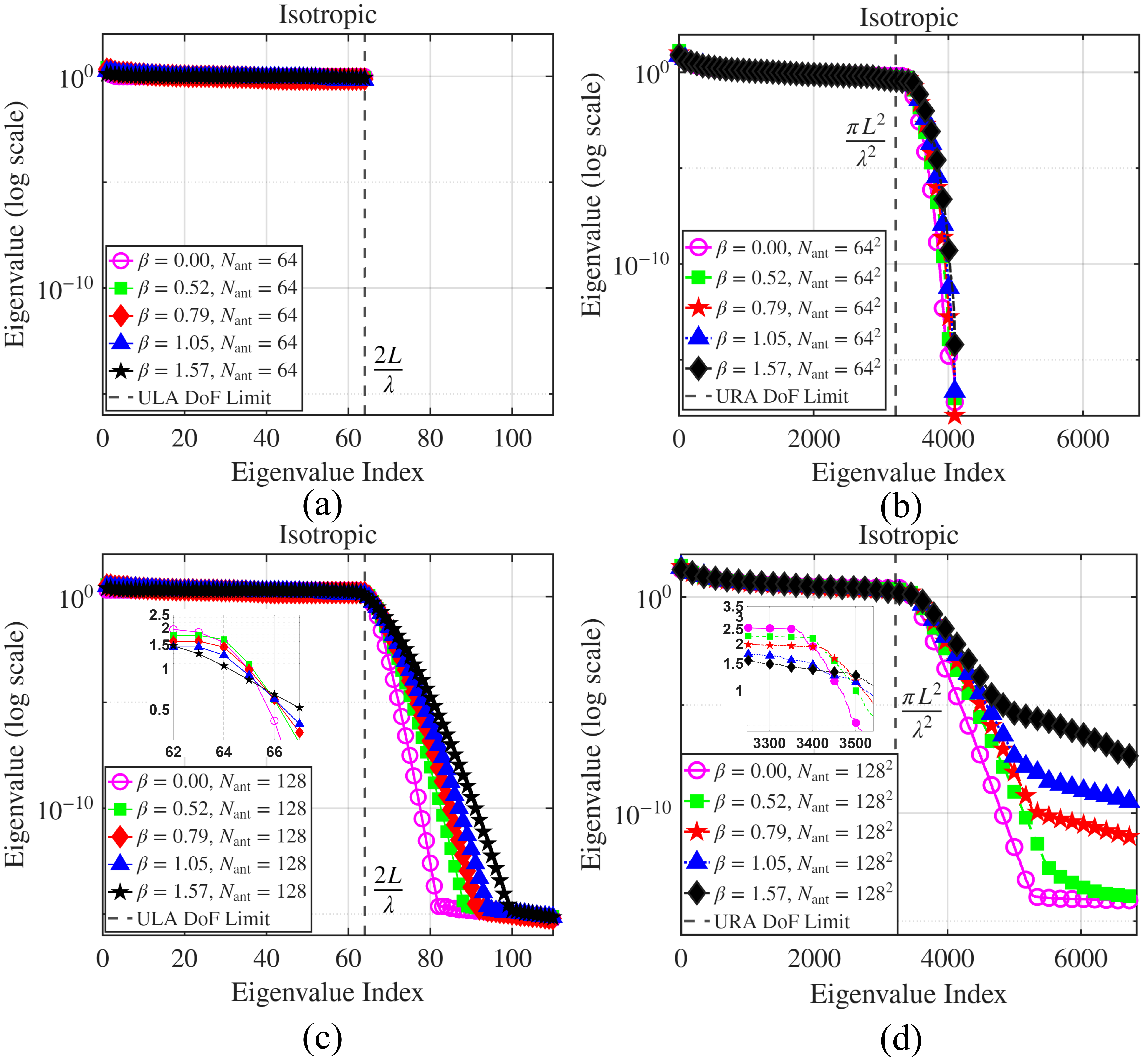}
        \captionsetup{font=footnotesize, labelsep=period}
\caption{Sorted eigenvalue spectra of the spatial correlation matrix for HoloCuRA under isotropic scattering. (a),(c) 1D HoloCuRA with $N_{\mathrm{ant}}=64$ and $N_{\mathrm{ant}}=128$, respectively; (b),(d) 2D HoloCuRA with $N_{\mathrm{ant}}=64^2$ and $N_{\mathrm{ant}}=128^2$, respectively. Eigenvalues are ordered in decreasing magnitude and plotted versus the (dimensionless) eigenvalue index on a logarithmic scale. Marker curves correspond to the curvature parameter $\beta$ (see legends). Vertical dashed lines indicate the spatial DoF limits, $2L/\lambda$ for the ULA and $\pi L^2/\lambda^2$ for the URA; insets in (c) and (d) zoom into the cutoff region.}
\label{fig_dofiso}
	\end{centering}
\end{figure}

\subsection{Spatial DoF Under Isotropic Scattering}

Building on the same coverage-extended isotropic correlation model, we next examine the effective DoF through the eigenvalue spectrum. As $\beta$ increases, the visible angular domain expands and the set of supported inter-element phase differences becomes richer. Consequently, spatial correlation weakens and spectral energy spreads from a few dominant modes to higher orders, which appears as a rightward shift of the spectral knee in Fig.~\ref{fig_dofiso}(a)--(b). Because the matrices have already been diagonal-normalized, this shift reflects a redistribution of modal energy rather than an overall power increase.

The detailed tail behavior is then set by aperture dimensionality. For 1D HoloCuRA, the additional support introduced by curvature remains relatively well aligned with the baseline one-dimensional manifold, so the energy redistribution is concentrated mainly among the leading eigenmodes and the high-order tail remains comparatively weak. For 2D HoloCuRA, by contrast, the extra aperture dimension allows the additional support to populate eigen-directions that are less aligned with the baseline subspace, which produces a more pronounced high-order tail.

The comparison between $\lambda/2$ and $\lambda/4$ further clarifies the role of holographic sampling. In 1D HoloCuRA, the $\lambda/2$ case is effectively close to critical sampling relative to the aperture DoF, so the spectrum appears as a near-flat plateau followed by a sharp drop. The $\lambda/4$ case does not increase the physical DoF, but resolves the post-cutoff transition much more clearly, as highlighted by the inset in Fig.~\ref{fig_dofiso}(c). The same effect is even more visible in 2D HoloCuRA, where the denser sampling produces a much richer spectral trace around the cutoff in Fig.~\ref{fig_dofiso}(d). Across all three environments, this role of dense sampling is most transparent under isotropic scattering, because the channel excites the full set of aperture-supported modes and the continuous-aperture interpretation is least obscured by directional sparsity.

\subsection{Joint Isotropic Insights on Correlation, SnS, and DoF}

This isotropic case plays a distinct role in the paper. It is not merely a third propagation scenario, but the continuous-aperture reference against which the stronger and more structured non-stationarity of the LoS and NLoS cases should be understood. The closed-form correlation kernels show explicitly how curvature enters through geometry and coverage extension, while the SnS results show that this geometric effect alone produces only weak array-domain heterogeneity under isotropic averaging.
At the same time, the DoF results clarify the main contribution of the holographic viewpoint. Dense sampling does not create new physical modes for a fixed aperture; instead, it exposes the underlying continuous-aperture eigenspectrum more faithfully and makes curvature-dependent redistribution of modal energy analytically visible. In this sense, isotropic scattering provides the cleanest link among correlation, SnS, DoF, and holographic sampling, and therefore serves as the theoretical anchor of the full HoloCuRA characterization developed in this paper.

\section{Realizable Antenna Port Modes: Bridging SnS and Physical DoF}

\subsection{Motivation and Definition}

Having characterized array-domain SnS via $d_{\mathrm{PoVi}}$ heatmaps and physical spatial DoF via eigen-spectra in the preceding sections, we now connect these two views at the implementation level for port-limited HoloCuRA.
We focus on the LoS case, where low-altitude links are often LoS-dominant and SnS is primarily geometry driven by spherical-wave gradients, curvature projection, and VR gating, making the resulting port mode budget physically interpretable and transferable.
While physical DoF specifies the modal ceiling of the full aperture, practical HoloCuRA implementations are realized through a finite number of subarray ports.
This motivates the following question: under LoS, how many subarray ports are effectively non-redundant and can therefore act as distinct aperture modes?
\begin{figure}[htpb]
	\centering
	\includegraphics[width=0.485\textwidth]{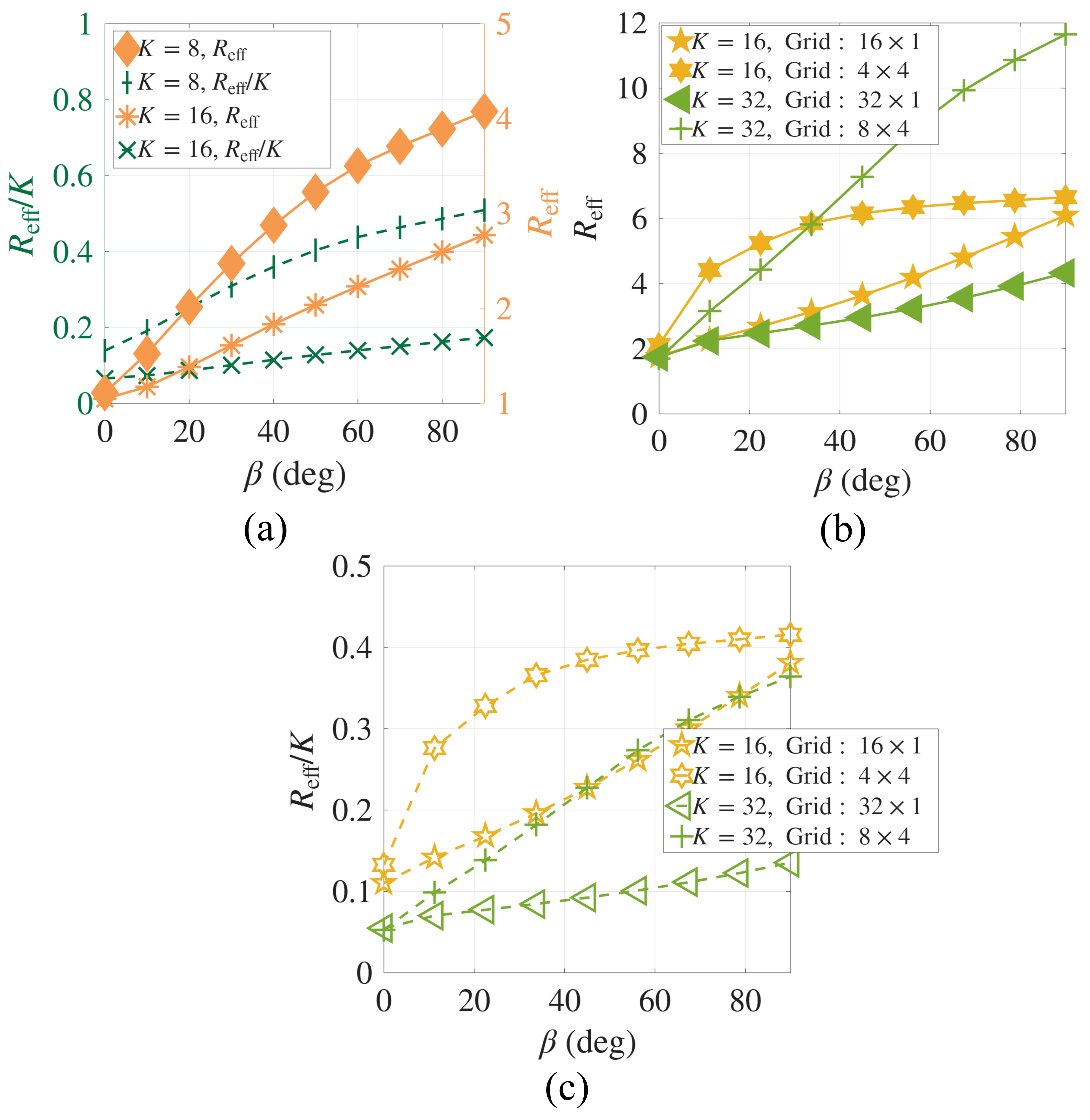}
    \captionsetup{font=footnotesize, labelsep=period}
	\caption{Realizable port mode versus $\beta$ for HoloCuRA under low-altitude half-space averaging. A near-field LoS channel is considered with $f_c=30~\mathrm{GHz}$ and fixed user range $r=2~\mathrm{m}$.
(a) 1D HoloCuRA ($L=0.32~\mathrm{m}$, $N=128$): $R_{\mathrm{eff}}(r,\beta)$ (right axis) and $R_{\mathrm{eff}}/K$ (left axis) for $K\in\{8,16\}$.
(b) 2D HoloCuRA ($0.32~\mathrm{m}\times 0.32~\mathrm{m}$, $128\times 128$): $R_{\mathrm{eff}}(r,\beta)$ for $K\in\{16,32\}$ under representative $z$-cut and 2D grid port arrangements.
(c) $R_{\mathrm{eff}}/K$ for the same arrangements as in (b).
For comparison, the 2D arrangements are as follows. For $K=16$: Grid $16\times 1$ (tile $8\times 128$ elements) versus Grid $4\times 4$ (tile $32\times 32$). For $K=32$: Grid $32\times 1$ (tile $4\times 128$) versus Grid $8\times 4$ (tile $\approx 16\times 32$).}
	\label{fig_port}
\end{figure}
The proposed PoVi-CMD provides a bounded pairwise dissimilarity that jointly captures VR gating, power imbalance, and correlation-shape mismatch across subarrays.
We therefore use it to construct an inter-port similarity matrix and a realizable port mode dimension.

For a HoloCuRA implemented with $K$ subarray ports, define at each
$\Omega=(\theta,\phi)$
\begin{equation}
[\mathbf{S}(\Omega)]_{k\ell} \triangleq s_{k\ell}(\Omega)=
\begin{cases}
1, & k=\ell,\\[1mm]
\sqrt{1-d_{\mathrm{PoVi}}^{k\ell}(\Omega)}, & k\neq \ell,
\end{cases}
\label{eq:S_def}
\end{equation}
where $d_{\mathrm{PoVi}}^{k\ell}(\Omega)$ is defined in \eqref{eq:dPoVi}, with $\gamma_{k\ell}(\Omega)$ and $\alpha_{k\ell}(\Omega)$ given in \eqref{eq:gamma}--\eqref{eq:alpha}, and $f_q(\cdot)$ in \eqref{eq:fp}.
Larger PoVi-CMD values imply weaker inter-port similarity and hence lower port redundancy.
Here, $\mathbf{S}(\Omega)$ is used only as a bounded symmetric summary of inter-port similarity under LoS; it is not an element-level covariance matrix, and the construction below does not require a PSD or kernel interpretation.

We then define a realizable port mode dimension in the same trace-ratio form used for physical DoF:
\begin{equation}
r_{\rm eff}(\Omega)
=
\frac{\big(\mathrm{tr}\,\mathbf{S}(\Omega)\big)^2}{\|\mathbf{S}(\Omega)\|_F^{2}}
=
\frac{\big(\mathrm{tr}\,\mathbf{S}(\Omega)\big)^2}{\mathrm{tr}\big(\mathbf{S}^2(\Omega)\big)}.
\label{eq:reff_def}
\end{equation}

Since $\mathrm{tr}\,\mathbf{S}(\Omega)=K$,
\begin{equation}
r_{\rm eff}(\Omega)
=
\frac{K^2}{K+2\sum_{k<\ell}s^2_{k\ell}(\Omega)}
=
\frac{K^2}{K+2\sum_{k<\ell}\!\big(1-d_{\mathrm{PoVi}}^{k\ell}(\Omega)\big)}.
\label{eq:reff_closed}
\end{equation}

Moreover, since $0\le s_{k\ell}(\Omega)\le 1$ for all $k\neq \ell$, we have
\begin{equation}
K\le \mathrm{tr}\!\big(\mathbf{S}^2(\Omega)\big)
=K+2\sum_{k<\ell}s_{k\ell}^2(\Omega)\le K^2,
\label{eq:trace_bound}
\end{equation}
which together with \eqref{eq:reff_def} yields
\begin{equation}
1\le r_{\rm eff}(\Omega)\le K.
\label{eq:reff_bound}
\end{equation}

Hence, $r_{\rm eff}(\Omega)\to 1$ when most ports remain highly similar, whereas $r_{\rm eff}(\Omega)\to K$ when SnS strongly differentiates the ports.
Thus, $r_{\rm eff}(\Omega)$ quantifies the number of approximately non-redundant port modes available at $\Omega$.

\subsection{Averaging and Relation to Physical DoF}

A scenario-level budget is obtained by averaging over the operating domain $\Omega_{\rm dom}$.
For low-altitude LoS operation, $\Omega_{\rm dom}$ is taken as the relevant half-space:
\begin{equation}
R_{\rm eff}(r,\beta)=\mathbb E_{\Omega\in\Omega_{\rm dom}}\!\big[r_{\rm eff}(\Omega)\big].
\label{eq:avg_reff}
\end{equation}
Unless otherwise stated, the expectation in \eqref{eq:avg_reff} is taken with respect to a uniform angular measure over the low-altitude half-space.
The quantity $R_{\rm eff}(r,\beta)$ is not a new physical DoF.
Rather, $\mathrm{DoF}_{\rm phys}$ remains the modal ceiling of the full aperture, whereas $R_{\rm eff}$ describes how many approximately independent port modes can be realized under a given subarray-port arrangement.
This motivates the practical bound
\begin{equation}
N_{\rm port\text{-}modes}(r,\beta)\ \lesssim\ 
\min\!\big\{R_{\rm eff}(r,\beta),\,\mathrm{DoF}_{\rm phys},\,N_{\rm RF}\big\}.
\label{eq:portmode_bound}
\end{equation}
where $N_{\rm RF}$  denotes the number of RF chains.
\subsection{Impact of Curvature and Port Arrangement for the Low-Altitude}

We evaluate $R_{\mathrm{eff}}(r,\beta)$ over the low-altitude half-space using the same setup as in Sec.~IV, with subarray sizes chosen to satisfy the local-stationarity condition.
The local-stationarity condition is used only to define an admissible port granularity, whereas $R_{\mathrm{eff}}$ summarizes the residual inter-port similarity across such admissible ports.

For 1D HoloCuRA, Fig.~\ref{fig_port}(a) shows that both $R_{\mathrm{eff}}$ and $R_{\mathrm{eff}}/K$ increase monotonically with $\beta$, indicating that curvature reduces inter-port similarity and enlarges the realizable port mode budget.
However, increasing the number of subarray ports does not necessarily increase $R_{\mathrm{eff}}$: relative to $K=8$, the $K=16$ case yields smaller $R_{\mathrm{eff}}$ and lower $R_{\mathrm{eff}}/K$, revealing substantial redundancy under excessive subdivision in 1D.
A similar redundancy effect appears for 2D HoloCuRA under $z$-cut arrangements in Fig.~\ref{fig_port}(b)--(c): $R_{\mathrm{eff}}$ grows with $\beta$ but decreases when $K$ increases from 16 to 32 because the additional ports remain highly similar.
By contrast, the 2D grid arrangements markedly improve both $R_{\mathrm{eff}}$ and $R_{\mathrm{eff}}/K$, indicating that localized 2D port organization better captures incident-field diversity and reduces similarity accumulation.

Collectively, these trends directly yield low-altitude design guidance.
When $\beta$ is small or the chosen arrangement gives a small $R_{\mathrm{eff}}$, subarray merging is preferred for robust coherent focusing.
Conversely, when $\beta$ is sufficiently large and the arrangement achieves a larger $R_{\mathrm{eff}}$, separated ports can support more agile beamforming, multiplexing, or nulling, subject to \eqref{eq:portmode_bound}.
Hence, $R_{\mathrm{eff}}(r,\beta)$ serves as an implementation-level budget for port-limited HoloCuRA, while $\mathrm{DoF}_{\rm phys}$ remains the physical modal ceiling.

\section{Conclusion}

In this paper, we developed a spatial characterization framework for low-altitude HoloCuRA by jointly analyzing array-domain SnS and spatial DoF under LoS, 3GPP NLoS, and isotropic scattering propagation environments. For SnS, we proposed PoVi-CMD and established a two-stage local-to-global procedure, in which VR-aware local admissibility first determines physically meaningful subarray partitions and full-aperture SnS is then characterized on that basis. The results show that LoS SnS is driven jointly by spherical-wave gradients, curvature-dependent projection, and self-occlusion propagation environments, whereas in NLoS curvature reshapes non-isotropic angular support.

For spatial DoF, we adopted the R\'enyi-2 effective rank and clarified the role of holographic sampling through a unified propagation-correlation view. In particular, the isotropic case serves as a theoretical anchor: 1D and 2D HoloCuRAs admit tractable closed-form correlation expressions, which show that the effective spatial DoF is governed mainly by aperture dimensionality and propagation support, while curvature acts primarily through support reshaping and tail-level eigen-spectrum perturbation. Dense holographic sampling does not create new physical modes for a fixed aperture size, but suppresses discretization artifacts and reveals the underlying continuous-aperture correlation and eigenspectrum more faithfully. Finally, by mapping PoVi-CMD into a port-similarity matrix, we connected SnS to a realizable antenna port mode budget, thereby bridging spatial characterization and port-limited low-altitude holographic aperture design.

\appendices

\section{Derivations for 1D HoloCuRA Correlation With Coverage Extension}
\label{app:Flexible_ula_deriv}

\subsection{Coverage extension term and 1D integral form}

With $\Omega_{\mathrm{ext}}(\beta)$ in \eqref{eq:Omega_ext_ula_main},
\begin{equation}
\label{eq:app_ula_Rextra_def}
R_{\mathrm{extra},m,n}(\beta)=
\int_{\Omega_{\mathrm{ext}}(\beta)}
\frac{\sin\theta}{2\pi}\,
e^{j\Delta_{m,n}(\theta,\varphi)}\,d\varphi\,d\theta .
\end{equation}
For a given $\theta$, define
\begin{equation}
\label{eq:app_ula_Iphi_def}
I_{\varphi}^{(\pi,2\pi)}(\theta)=
\int_{\pi}^{2\pi} e^{j\Delta_{m,n}(\theta,\varphi)}\,d\varphi .
\end{equation}
Let $z \triangleq b\sin\theta\sin c$ (with $b,c$ in \eqref{b,c}). Then
\begin{equation}
\label{eq:app_ula_Iphi_factor}
I_{\varphi}^{(\pi,2\pi)}(\theta)=
e^{jb\cos\theta\cos c}
\int_{\pi}^{2\pi} e^{-jz\sin\varphi}\,d\varphi .
\end{equation}

\noindent\textbf{Exact form.}
With $\varphi'=\varphi-\pi$ and $\sin(\varphi'+\pi)=-\sin\varphi'$~\cite{Bowman2012Bessel},
\begin{equation}
\label{eq:app_ula_phi_exact}
\begin{aligned}
\int_{\pi}^{2\pi} e^{-jz\sin\varphi}\,d\varphi
&=\int_{0}^{\pi} e^{jz\sin\varphi'}\,d\varphi' \\
&=\int_{0}^{\pi}\cos(z\sin\varphi')\,d\varphi' \\
&\quad +j\int_{0}^{\pi}\sin(z\sin\varphi')\,d\varphi' \\
&= \pi J_0(z) + j\,\Xi(z).
\end{aligned}
\end{equation}
where $\Xi(z)\triangleq\int_{0}^{\pi}\sin(z\sin t)\,dt$.

\noindent\textbf{Even-part (zeroth-order) approximation.}
Retaining only the real (even) part of \eqref{eq:app_ula_phi_exact},
\begin{equation}
\label{eq:app_ula_phi_even}
\int_{\pi}^{2\pi} e^{-jz\sin\varphi}\,d\varphi
\approx \pi J_0(z).
\end{equation}
Thus,
\begin{equation}
\label{eq:app_ula_Iphi_0}
I_{\varphi}^{(\pi,2\pi)}(\theta)\approx
\pi e^{jb\cos\theta\cos c}\,
J_0\!\big(b\sin\theta\sin c\big).
\end{equation}
Substituting \eqref{eq:app_ula_Iphi_0} into \eqref{eq:app_ula_Rextra_def} yields
\begin{IEEEeqnarray}{rCl}
R_{\mathrm{extra},m,n}^{(0)}(\beta)
&=& \frac12\Big(\int_{0}^{\beta}+\int_{\pi-\beta}^{\pi}\Big)\nonumber\\
&& \sin\theta\,
e^{j b\cos\theta\cos c}\,
J_0\!\big(b\sin\theta\sin c\big)\,d\theta .
\label{eq:app_ula_Rextra_theta}
\end{IEEEeqnarray}
Using the cap symmetry $\theta\mapsto\pi-\theta$ gives
\begin{equation}
\label{eq:app_ula_Rextra_cos}
\begin{aligned}
R_{\mathrm{extra},m,n}^{(0)}(\beta)
&=\int_{0}^{\beta}\sin\theta\,
\cos\!\big(b\cos\theta\cos c\big)\,
J_0\!\big(b\sin\theta\sin c\big)\,d\theta .
\end{aligned}
\end{equation}
Finally, with $u=\cos\theta$, and $b, c$ in \eqref{b,c}
\begin{equation}
\label{eq:app_ula_Rextra_1d}
R_{\mathrm{extra},m,n}^{(0)}(\beta)
=\int_{\cos\beta}^{1}
\cos\!\big(bu\cos c\big)\,
J_0\!\Big(b\sqrt{1-u^{2}}\sin c\Big)\,du .
\end{equation}

\section{Derivations for 2D HoloCuRA Correlation With Coverage Extension}
\label{app:Flexible_ura_deriv}

\subsection{Coverage extension term and 1D integral form}

The phase difference between $(m,n)$ and $(m',n')$ is~\cite{VTC2025}
\begin{equation}
\label{eq:app_ura_phase_sep}
\Delta_{m,n}-\Delta_{m',n'}
=A(\theta)\cos\varphi+B(\theta)\sin\varphi+C(\theta).
\end{equation}
Define
\begin{equation}
\label{eq:app_ura_short}
t_{nn'}\triangleq \beta-\frac{\psi_n+\psi_{n'}}{2},
\qquad
s_{nn'}\triangleq \sin\!\Big(\frac{\psi_n-\psi_{n'}}{2}\Big).
\end{equation}
Then
\begin{equation}
\label{eq:app_ura_ABC}
\begin{aligned}
A(\theta)&=(m-m')d_x\sin\theta, \\
B(\theta)&=-2R\sin\theta\;\sin(t_{nn'})\,s_{nn'}, \\
C(\theta)&=\;\;2R\cos\theta\;\cos(t_{nn'})\,s_{nn'}.
\end{aligned}
\end{equation}
Let $k_0=2\pi/\lambda$ and define
\begin{equation}
\Phi(\theta,\varphi)\triangleq
A(\theta)\cos\varphi+B(\theta)\sin\varphi+C(\theta).
\end{equation}
With $\Omega_{\mathrm{ext}}(\beta)$ in \eqref{eq:Omega_ext_ula_main},
\begin{equation}
\label{eq:app_ura_Rextra_def}
R_{\mathrm{extra}}(\beta)=
\int_{\Omega_{\mathrm{ext}}(\beta)}
\frac{\sin\theta}{2\pi}\,
e^{-jk_0\Phi(\theta,\varphi)}\,d\varphi\,d\theta .
\end{equation}
For a given $\theta$, define
\begin{equation}
\label{eq:app_ura_Iphi_def}
I_{\varphi}^{(\pi,2\pi)}(\theta)\triangleq
\int_{\pi}^{2\pi} e^{-jk_0\Phi(\theta,\varphi)}\,d\varphi .
\end{equation}
Let $\rho(\theta)\triangleq\sqrt{A(\theta)^2+B(\theta)^2}$ and choose $\delta(\theta)$ such that
\begin{equation}
\cos\delta(\theta)=\frac{A(\theta)}{\rho(\theta)},
\qquad
\sin\delta(\theta)=\frac{B(\theta)}{\rho(\theta)}.
\end{equation}
Then $A\cos\varphi+B\sin\varphi=\rho\cos(\varphi-\delta)$ and
\begin{equation}
\label{eq:app_ura_Iphi_chain}
\begin{aligned}
I_{\varphi}^{(\pi,2\pi)}(\theta)
&=e^{-jk_0C(\theta)}
\int_{\pi}^{2\pi}
e^{-jk_0\rho(\theta)\cos(\varphi-\delta(\theta))}\,d\varphi \\
&\overset{\varphi'=\varphi-\pi}{=}
e^{-jk_0C(\theta)}
\int_{0}^{\pi}
e^{+jk_0\rho(\theta)\cos(\varphi'-\delta(\theta))}\,d\varphi'.
\end{aligned}
\end{equation}

\noindent\textbf{Even/odd decomposition and zeroth-order approximation.}
For brevity, let
$x\triangleq k_0\rho(\theta)$ and $\delta\triangleq\delta(\theta)$.
Using Euler's identity,
\begin{equation}
\label{eq:app_ura_split}
\begin{aligned}
\int_{0}^{\pi} e^{jx\cos(\varphi'-\delta)}\,d\varphi'
&=\int_{0}^{\pi}\cos\!\big(x\cos(\varphi'-\delta)\big)\,d\varphi' \\
&\quad +j\int_{0}^{\pi}\sin\!\big(x\cos(\varphi'-\delta)\big)\,d\varphi'.
\end{aligned}
\end{equation}
The even (real) part admits the standard Bessel representation
$\int_{0}^{\pi}\cos(x\cos t)\,dt=\pi J_0(x)$.
Retaining only this even part yields
\begin{equation}
\label{eq:app_ura_phi_even}
\int_{0}^{\pi} e^{jx\cos(\varphi'-\delta)}\,d\varphi'
\approx \pi J_0(x).
\end{equation}
Thus,
\begin{equation}
\label{eq:app_ura_Iphi_0}
I_{\varphi}^{(\pi,2\pi)}(\theta)\approx
\pi e^{-jk_0C(\theta)}\,J_0\!\big(k_0\rho(\theta)\big).
\end{equation}
From \eqref{eq:app_ura_ABC}, $\rho(\theta)=D\sin\theta$ and $C(\theta)=E\cos\theta$ hold with
\begin{equation}
\label{eq:app_ura_DE}
\begin{aligned}
D&=\sqrt{(m-m')^2 d_x^2 + 4R^2 s_{nn'}^2 \sin^2(t_{nn'})},\\
E&=2R\,s_{nn'}\cos(t_{nn'}).
\end{aligned}
\end{equation}
Substituting \eqref{eq:app_ura_Iphi_0} into \eqref{eq:app_ura_Rextra_def} yields
\begin{IEEEeqnarray}{rCl}
R_{\mathrm{extra}}^{(0)}(\beta)
&=& \frac12\Big(\int_{0}^{\beta}+\int_{\pi-\beta}^{\pi}\Big)\nonumber\\
&& \sin\theta\,e^{-j k_0E\cos\theta}\,
J_0\!\big(k_0D\sin\theta\big)\,d\theta \nonumber\\
&=& \int_{0}^{\beta}\sin\theta\,
\cos\!\big(k_0E\cos\theta\big)\,
J_0\!\big(k_0D\sin\theta\big)\,d\theta \nonumber\\
&\stackrel{u=\cos\theta}{=}&
\int_{\cos\beta}^{1}\cos(k_0Eu)\,
J_0\!\big(k_0D\sqrt{1-u^2}\big)\,du .
\label{eq:app_ura_extra_1d}
\end{IEEEeqnarray}

\bstctlcite{IEEEexample:BSTcontrol}
\bibliographystyle{IEEEtran}
\bibliography{ref5}




\end{document}